%% file: main.tex
\documentclass[pdflatex, sn-mathphys-num]{sn-jnl}
\usepackage[utf8]{inputenc}

\usepackage{adjustbox}

\usepackage{tabulary}
\usepackage{multirow}%
\usepackage{amsmath,amssymb,amsfonts}%
\usepackage{amsthm}%
\usepackage{mathrsfs}%
\usepackage[title]{appendix}%
\usepackage[table,xcdraw]{xcolor}
\usepackage[comments]{exsitu}

\usepackage{textcomp}%
\usepackage{manyfoot}%
\usepackage{booktabs}%
\usepackage{algorithm}%
\usepackage{algorithmicx}%
\usepackage{algpseudocode}%
\usepackage{listings}%
\usepackage{graphicx}
\usepackage{subcaption} 
\usepackage{pdflscape}
\usepackage{ulem}

\begin{document}

\title[What's in a BIP? Exploring the Lived Experiences of Breaks In Presence]{What's in a BIP? Exploring the Lived Experiences of Breaks In Presence}

\author*[1]{\fnm{Jean-Philippe} \sur{Rivière}}\email{riviere-jp@univ-nantes.fr}

\author[2]{\fnm{Roman} \sur{Malo}}\email{roman.malo@univ-nantes.fr}

\author[1,2]{\fnm{Sarah} \sur{Varlin Grassi}}\email{sarah.varlin--grassi@etu.univ-nantes.fr}

\author[1]{\fnm{Yannick} \sur{Prié}}\email{yannick.prie@univ-nantes.fr}

\affil*[1]{\orgdiv{Nantes Université, École Centrale Nantes, CNRS}, \orgname{LS2N, UMR 6004}, \orgaddress{\city{Nantes}, \postcode{F-44000}, \country{France}}}

\affil*[2]{\orgdiv{Nantes Université, Laboratoire de psychologie des Pays de la Loire}, \orgname{LPPL, UR 4638}, \orgaddress{\city{Nantes}, \postcode{F-44000}, \country{France}}}

\abstract{Occasionally, individuals immersed in a Virtual Reality (VR) environment may experience distractions that disrupt their sense of presence, a phenomenon referred to as a break in presence (BIP). Better understanding BIPs is crucial to designing VR applications that keep their users present. BIPs have been studied using a variety of methods, exploring their origins or trying to detect them from physiological or behavioral measurements. However, despite the importance of understanding how they are actually lived and managed by VR users, very few studies focused on their phenomenological characterization. We employed micro-phenomenology to collect the descriptions of BIPs experienced by users (\textit{n}=14) of a height exposure VR application. We precisely modeled 57 BIP episodes, bringing to light a variety of experiences and behaviors. Four generic diachronic patterns of BIP episodes emerge: \textit{reflected-upon}, \textit{discarded}, \textit{self-preservation}, and \textit{contradictory mediation} BIPs. We discuss these in light of the PI/Psi model of presence, propose an awareness-based definition of BIPs, as well as three BIP-related design opportunities.}


\maketitle

\section{Introduction}

Central to the experience of Virtual Reality (VR) is presence, the sense of ``being there'' in a virtual world~\cite{witmer_measuring_1998}, accompanied by the illusion that the events unfolding are actually happening~\cite{slater_place_2009}. 
At times, presence can be disrupted, leading to interruptions, referred to in the literature as Breaks in Presence (BIPs)~\cite{slater_virtual_2000}. BIPs can have multiple causes, such as real-world distractions (a ringing phone in reality), the lack of realism or plausibility of VR environments, etc. They are usually seen as a negative aspect of VR experiences, as they are illusion breakers~\cite{brooks_1999_what}, and distract the users from whatever immersive activity they were conducting. Understanding what kinds of BIPs can occur~\cite{Pouke_2022_qualitative}, what are their causes~\cite{chertoff_improving_2008}, and automatically detecting their occurrences~\cite{gaspari_detection_2023} is therefore important to provide designers with guidelines on how to avoid them when possible, and how to manage them gracefully. 

BIPs are mostly considered as the absence of presence or as breaks appearing at specific moments, but there are surprisingly few studies to assess what they are from the perspective of the users who experience them. Though we know BIPs seem to be continuous~\cite{garau_temporal_2008}, vary in intensity~\cite{waterworth_focus_2001, chertoff_improving_2008}, and are managed differently by users~\cite{Pouke_2022_qualitative}, our knowledge of how this exactly happens remains limited. It seems then crucial to empirically explore the \textit{lived experiences} of BIPs (here we define lived experience as ``\textit{what a singular subject experiences at a given moment and in a given place: what he accesses in 'first person'}''~\cite[p.~15]{depraz_becoming_2003}). We want to understand how BIPs appear, unfold over time, and if and how they differ, from a user perspective.
From a design point of view, such insights can help develop strategies to support recovery from BIPs or diminish their impact on users.

Our main research goal is therefore to characterize the lived experiences of breaks in presence. We followed an approach based on phenomenology, a philosophical current that originated as a rigorous and systematic study of consciousness in the early 20th century through the work of Edmund Husserl. Phenomenology attempts to study the structures of consciousness as experienced from a first-person perspective~\cite{husserl1927phenomenology}, and gave rise to methods specifically designed to study lived experiences~\cite{Lumma2021, hoffding2023working}, called first-person methods\footnote{Authors diverge in their use of the terms first- and second-person methods: some classify interview-based techniques as second-person, reserving 'first-person' for approaches like introspection and auto-ethnography~\cite{hoffding2023working}, while others regard interviews that elicit lived experiences as first-person~\cite{Lumma2021}. In this paper, we adopt the latter view.}. With the advent of the third wave of Human-Computer Interaction~\cite{bodker2006second}, these methods have proven effective for examining the multifaceted nature of human experience in digital interaction—encompassing perceptual, aesthetic, affective, embodied, and cognitive dimensions~\cite{prpa_articulating_2020}—and are now widely adopted.

In this paper, we present an exploratory experimental study with 14 participants placed in a VR environment designed to induce presence through the fear of heights. We triggered various types of distractors (visual, haptic, and audio) during VR sessions to induce BIPs.
We then collected participants' lived experiences of BIPs using a first-person method called micro-phenome\-nological interview. We analyzed and modeled the phenomenological content of each BIP that was reported by the participants, and compared these BIP \textit{episodes} to find recurring BIP patterns. Our study is the first of its kind to investigate BIP from a first-person perspective, aiming to deepen the understanding of the phenomenon and to lay the groundwork for precise and systematic descriptions.

\section{Related Work}

We first review some works about presence, before focusing on breaks in presence and associated studies. We then present the fundamentals of the micro-phenomenological approach and how it has been used for studying lived experiences in VR.

\subsection{Presence and breaks in presence}

BIPs depend on presence. Presence is considered as a psychological state ---an inherently subjective internal experience--- that is not tied to any specific technological domain~\cite{lee_presence_2004}. For example, presence can be discussed in relation to VR, but also in other technology-mediated contexts, such as reading a book or watching a movie. In this paper, we specifically focus on presence in a VR environment.
To this day, there is still no consensus regarding the dimensions that constitute presence in VR~\cite{Felton2021} and multiple theoretical models exist. 

\subsubsection{Presence}
\label{sec:presence}

In their extended survey, \citeauthor{skarbez_survey_2017} categorized existing definitions of presence into three main categories\footnote{
It should be noted that these categories are not mutually exclusive. For example, some researchers argue that the non-mediation aspect of VR may actually result from the experience of spatial presence~\cite{Wirth_process_2007}.
}: \textit{being there}, \textit{non-mediation}, and \textit{others}~\cite{skarbez_survey_2017}. The \textit{being there} category encompasses most definitions, framing it as the feeling of being situated within an environment, often referred to as \textit{spatial presence}. Though some definitions do not explicitly address the interactional aspect of being in VR, many of them require users to feel actively engaged in the virtual world, reflecting a ``Gibsonian''\footnote{From the ecological psychologist James Gibson who insisted on the tight coupling between perception and action, and the direct perception of \textit{affordances} as possibilities for action offered by the environment.} approach to presence~\cite{steuer_defining_1992, skarbez_survey_2017}. For example, for \citeauthor{witmer_factor_2005}, presence is the ``psychological state of `\textit{being there}' mediated by an environment that engages our senses, captures our attention, and fosters our active involvement''\cite{witmer_factor_2005}. 

The second category encompasses definitions based on non-mediation, and conceptualizes presence as the user’s lack of attention to the technology. This can be because the focus is on users' thoughts (as in the ``suspension of disbelief''); or because they respond as if the medium was not present, e.g. when \citeauthor{lombard_at_1997} define presence as ``the perceptual illusion of non-mediation''~\cite{lombard_at_1997}. 

The last category regroups more marginal definitions. \citeauthor{lee_presence_2004} puts for instance the accent on the user experiencing virtual objects as real when present in VR~\cite{lee_presence_2004}. \citeauthor{waterworth_focus_2001} take a different angle by defining presence as a function of attention, proposing three dimensions for it: locus (whether attention is directed toward the real or virtual world), focus (whether information is processed perceptually or conceptually), and sensus (whether the user is conscious or unconscious of said world)~\cite{waterworth_focus_2001}. Interestingly, as noted by~\citeauthor{murphy_what_2020}, attention seems closely tied to presence. Slater argued that attention is not strictly required for spatial presence: even if our minds wander, our perception of where we are does not necessarily change~\cite{slater_separate_2022}. It also remains unclear whether a lack of attention to the VR world should be considered a break. The relationship between presence and attention is still insufficiently understood and calls for further clarification~\cite{murphy_what_2020}.

\citeauthor{skarbez_survey_2017} also reviewed existing models of presence and proposed a framework, defining it as the combination of Place Illusion (PI), Plausibility Illusion (Psi), and Social Presence Illusion (SPI)~\cite{skarbez_survey_2017}. PI and Psi were previously introduced by \citeauthor{slater_place_2009}: PI refers to the illusion of being physically present in a virtual environment, which is closely aligned with the ``being there'' conception of presence, while Psi is defined as ``the illusion that what is apparently happening is really happening''~\cite{slater_place_2009}. PI is primarily driven by the degree of immersion of the system (i.e., the objective characteristic of a virtual environment system~\cite{Slater_note_2003}) while Psi depends on the coherence of the virtual scenario. According to \citeauthor{skarbez_survey_2017}, presence naturally emerges when PI and Psi are jointly experienced in a virtual environment, consistent with Slater’s view~\cite{slater_place_2009}.
As for Social Presence Illusion (SPI), it refers to the illusion of ``being together with and engaging with a real sentient being''. \citeauthor{skarbez_survey_2017} consider it as equally important as PI and Psi in collaborative virtual environments. However, we consider SPI to be outside the scope of the present work, as it is specifically related to social VR.

Depending on the definition of presence one chooses, the associated definition of BIP can vary. A BIP could be understood as the feeling of not being in the VR environment, or a shift of attention toward the mediating technology. Other possibilities are related to the user experiencing objects as non-actual entities, or breaks in either the locus, the focus, or the sensus of attention. In this paper, we adopt the PI/Psi model of presence, which implies that either PI or Psi can be disrupted. Consequently, we will focus on designing distractors capable of breaking either PI or Psi.

\subsubsection{Breaks in presence}

The concept of BIP was introduced by~\citeauthor{slater_virtual_2000} as a way to measure presence. They defined a break as a moment when a participant ceases to respond to the virtual environment and instead reacts voluntarily or involuntarily to the sensory inputs from the real-world~\cite{slater_physiological_2003}. \citeauthor{Zahorik_presence_1998}, drawing on the philosopher Martin Heidegger, framed BIPs through their mediation-based definition of presence (``Presence is tantamount to successfully supported action in the environment''). In this view, the VR headset remains ``Ready-to-hand'' as long as users are present in the virtual world and unaware of the medium, but during a breakdown it becomes ``Present-at-hand'', as the medium suddenly reveals itself, forcing them into a reflective stance~\cite{Zahorik_presence_1998}. In the view of \citeauthor{Wirth_process_2007}, BIPs occur when spatial presence is lost, for example, when interaction with the environment is disrupted (e.g., due to a tracking failure)~\cite{Wirth_process_2007}. All these definitions of BIPs consider presence, and by extension breaks, in a binary mode: one is either present or absent to the virtual environment. 

However, Slater acknowledged in later works that such ``on/off'' thinking might oversimplify a more complex phenomenon. He suggested that an occurrence of a BIP could be better understood as a continuous phenomenon, with identifiable beginning and ending~\cite{slater_virtual_2000}. \citeauthor{spagnolli_immersionemersion_2002} investigated presence in VR during moments of breakdown and found that users could maintain a sense of presence in the virtual environment even while engaging with real-world stimuli, for instance, when interacting with an experimenter. They argue that describing disruption in VR as a sudden emersion is inadequate, suggesting instead that presence can be simultaneously distributed across both real and virtual settings~\cite{spagnolli_immersionemersion_2002}. 
An associated interesting perspective is offered by \citeauthor{murphy_what_2020}, who reject the strict on/off conception of presence. Instead, they use Biocca’s three-pole model~\cite{biocca2003can} to argue that the feeling of presence exists along a continuous spectrum, oscillating between three loci of presence: a physical space (reality), a virtual space, and a mental imagery space. An individual can shift from space to space (voluntarily or involuntarily), depending on idiosyncratic thresholds, the crossing of which would result in a BIP~\cite{murphy_what_2020}. They also acknowledge that an individual can feel present in multiple spaces simultaneously (hybrid presence).

Following these perspectives, we argue that BIPs are continuous and complex phenomena requiring investigation both into how they unfold over time~\cite{garau_temporal_2008} and how they feel to the users, which we aim to address in our study.

\subsubsection{Empirical studies of breaks in presence}

As for presence\footnote{Interested readers can refer to~\cite{Laarni2015,souza_measuring_2021} for details on the different methods for measuring presence.}, there are two main ways to measure BIPs: objective methods that use physiological or behavioral responses, and subjective methods based on questionnaires, self-reports and interviews.

\paragraph{Objective measures of BIPs}

Several correlations between presence and physiological measures have been identified in experimental settings, e.g. between reported levels of presence and increased heart rate~\cite{meehan_physiological_2002} or skin conductance\cite{peterson_effects_2018}\footnote{See~\cite{Felton2021} for a review}. However, most of these studies did not explicitly focus on detecting BIPs. Among those  doing so, \citeauthor{liebold_continuous_2017} induced BIPs into video games using distractors such as error messages, loss of character control, game crashes, floating objects, or the experimenter entering the room, showing the correlation of some of their occurrences with physiological and behavioral data (here heart rate, skin conductance, blinking, etc.)~\cite{liebold_continuous_2017}. Another study used psychophysiological (heart rate, pupil diameter, electrodermal activity) and behavioral (Secondary Task Reaction Times) measures to specifically detect BIPs induced with external sounds~\cite{Holl2025}. 
Lastly, EEG has also been shown to be effective in detecting BIPs in experimental settings~\cite{Savalle2024}.

\paragraph{Subjective measures of BIPs}

Questionnaires remain the most widely used method for assessing presence with subjective measures~\cite{schwind2019using}. However, they fall short when it comes to capturing BIPs. Although a low presence score may reflect one or several BIPs, one cannot assume that a BIP has actually occurred, and the underlying causes of a low score remain unclear. In fact, presence questionnaires provide little to no information about when (and if) breaks occurred, how they were experienced, or what triggered them. As \citeauthor{slater2004colorful} provocatively argued, while presence (and by extension BIPs) can likely be measured through questionnaires, gaining a deeper understanding of the phenomenon requires other approaches~\cite{slater2004colorful}.

Other self-report measures have been used to study BIPs. \citeauthor{slater_virtual_2000} had their participants perform a task in VR, and verbally report~\cite{slater_virtual_2000}, or press a button whenever they felt they had transitioned from the virtual to the real-world~\cite{slater_physiological_2003}. Continuous measures, e.g. the use of a slider, could also be used to investigate the intensity of presence over time. This is the case for \citeauthor{chung_temporal_2012} who induced BIPs on participants playing video games, showing that BIPs had varying impacts on the playing experience and that recovery time was not always constant~\cite{chung_temporal_2012}.

To our knowledge, only two studies used interviews to explicitly investigate BIPs empirically.
\citeauthor{garau_temporal_2008} generated four two-second screen whiteouts in a CAVE displaying a bar scene~\cite{garau_temporal_2008}. The thematic analysis of the debriefing interviews revealed that BIPs could vary in intensity and that users made efforts to recover from them. They also asked their participants to draw a graph representing their sense of presence over time, obtaining a general depiction of its fluctuations. On their side, \citeauthor{Pouke_2022_qualitative} asked twenty participants to describe BIPs they experienced during a VR game~\cite{Pouke_2022_qualitative}, identifying occurrences of BIPs as either breaks in PI or Psi. We can also mention a third study that provides valuable insights on BIPs, although it was not its explicit focus: \citeauthor{slater_sentiment_2023} studied participants’ experiences at a VR rock concert and identified and described several instances of breaks in Psi~\cite{slater_sentiment_2023}. 

As a last remark, all the studies we just mentioned provided examples of BIPs but did not analyze their temporal structures in details. The literature lacks when it comes to describing how BIPs unfold over time, or what characterizes them from an experiential point of view, a gap we address by proposing a detailed investigation into their phenomenology and temporal unfolding.

\subsubsection{Provoking BIPs}
\label{sec:interalExternalBIP}

There exist two main approaches to studying BIPs. One consists in placing participants in a VR environment and wait until BIPs occur naturally~\cite{Pouke_2022_qualitative, slater_virtual_2000, spagnolli_immersionemersion_2002}. 
The other involves using distractors to induce BIPs and study their occurrences~\cite{slater_physiological_2003, garau_temporal_2008, chung_2009_measuring, chung_temporal_2012, liebold_continuous_2017}. In this latter case, researchers design their own distractors. For example, \citeauthor{chung_temporal_2012} created technical anomalies that could appear in a real video-game, such as loss of frame rate, loss of audio, inverted control mapping, and black-out screen~\cite{chung_temporal_2012}, while \citeauthor{garau_temporal_2008} introduced blackouts in their CAVE environment~\cite{garau_temporal_2008}. 

On a more theoretical level, \citeauthor{chertoff_improving_2008} identified five types of distractors that can lead to a BIP. \textit{External distractors}, when the external world interferes with the VR experience (e.g., auditory disturbances such as a drilling sound from the street); \textit{Internal distractors}, when the users themselves are the source (e.g., when daydreaming); \textit{Inconsistent mediation}, when the environment fails to consistently support its output message or media (e.g., loss of frame rate); \textit{Contradictory mediation}, when the environment contradicts participants’ established schemas (e.g., a door that cannot be opened); and \textit{Unrefined mediation}, when the mediated environment triggers too many or contradictory schema (e.g., cognitive overload)~\cite{chertoff_improving_2008}. 

A central question remains: how should we design distractors? To date, no dedicated taxonomy exists to systematically guide the creation of distractors. Following our approach, we rely on the PI/Psi framework, which states that distractors should disrupt immersion or undermine the coherence of the virtual world. To operationalize this approach, we will build on \citeauthor{chertoff_improving_2008}'s types of distractors.

\subsection{The micro-phenomenological approach}

BIPs are complex phenomena that demand in-depth investigation through qualitative approaches to capture users’ lived experiences. However, lived experiences are often challenging to verbalize, as psychologist Pierre Vermersch observed~\cite{vermersch1994explicitation}, a  difficulty that extends to those related to BIPs. During traditional semi-structured interviews, participants tend to rationalize and interpret their experiences, often sharing judgments, beliefs, preferences, or theoretical knowledge instead of precisely describing their actual actions and sensations. Vermersch argued that while these analytical dimensions are important for understanding what happened, they do not constitute the experience itself~\cite{vermersch1994explicitation}. This rationalization occurs because a large part of our daily experience is not immediately accessible to reflective consciousness and verbal description~\cite{petitmengin_discovering_2019}. Many actions (e.g., opening a door) and perceptions (e.g., the color of the door) are experienced in a ``pre-reflective'' manner. 

Given this, Vermersch introduced a set of strategies and questions to guide interviewees in both recalling a past situation and verbalizing their experiences~\citep{cahour_analyzing_2016}. In such interviews, the interviewer seeks to induce an ``evocation state'' in which the interviewee re-lives an instance of the action under investigation and describes it. This state is identifiable through cues such as detached gaze, the use of first-person pronouns, or hand gestures. The interviewer’s goal is to help the interviewee remain in this state while gently probing what happened for them across various experiential dimensions, such as cognition, perception, emotion, sensation, etc. This set of techniques became the \textit{entretien d'explicitation} (in French), better known in English as the micro-phenomenological interview~\cite{petitmengin:hal-02524340}. Unlike classic semi-structured interviews, the interviewers do not rely on a predetermined set of questions. Instead, they focus on specific moments of past lived experience, following precise steps that require specialized training.

This type of interview is of utmost interest for our research question because it is specifically designed to capture the lived experience of internal phenomena, such as BIPs. Another key point is that contrary to other first-person methods such as thinking aloud~\cite{van1994think} or descriptive experience sampling~\cite{hurlburt_descriptive_2006}, that modify or interrupt the task~\citep{cahour_analyzing_2016}, it enables us to collect lived experiences afterwards, at a fine-grained level. Also, contrary to self-confrontation~\cite{theureau:hal-01107065}, it does not require traces of activity such as documents or videos. 

Moreover, micro-phenomenology has already been used to evaluate VR systems~\cite{quesnel_creating_2018, stepanova_understanding_2019, montuwy_using_2019, riviere_toward_2024}, and to better understand associated phenomena, such as breathing awareness~\cite{prpa_attending_2018}, the phenomenology of embodiment in VR~\cite{Hu2025}, or what happens when a user exits VR~\cite{knibbe_dream_2018}. In this last study, \citeauthor{knibbe_dream_2018} sought to better understand the specific moment when a user takes off a headset~\cite{knibbe_dream_2018}\footnote{Interestingly, they argue that this ``exit moment'' could be assimilated to a BIP, but since users do not continue using the VR system or attempt to return to the virtual environment, it also differs significantly from the BIPs we study.}. They reported intriguing findings: some participants mentally exited VR before physically removing the headset, while others only considered themselves out of VR upon regaining visual and auditory senses. 
They also described social experiences of being observed and sensory experiences of visually readjusting to the real environment. 

When it comes to analyzing micro-phenomenological data, two main methods stand out: thematic analysis or diachronic/synchronic
analysis. Both adopt a bottom-up approach, seeking to construct findings directly from participants' accounts. \textit{Thematic analysis}~\cite{braun_thematic_2022}, a classical qualitative method, has been largely used in the micro-phenomenological literature (e.g~\cite{hogan_elicitation_2016, knibbe_dream_2018}), however it tends to discard the temporal as well as the individual aspects of the lived experience, because the analyst considers the set of transcriptions as an indiscriminate pool of utterances from which to construct themes that apply to the whole corpus. 

\textit{Diachronic/synchronic analysis}~\cite{petitmengin_discovering_2019,valenzuela_procedure_2019}
has been proposed by the proponents of the micro-phenomenological approach. It is more and more used (see for instance~\cite{alcaraz_awareness_2021, riviere_toward_2024}), because it allows to make the most of the richness of the data collected by the interviews. There, the analyst tries to model precise episodes of lived experience from each interview before generalizing from these individual cases. Modeling is done along the \textit{diachronic} dimension, which describes the temporal organization of the experience in moments, and the \textit{synchronic} dimension, which focuses on the constituents of these moments of experience\footnote{The adjectives ``diachronic'' and ``synchronic'' belong to the vocabulary of the micro-phenomenological approach, we will use these two terms throughout the remainder of the paper to denote these aspects of lived experience.}.

This method, that we further detail in section~\ref{sec:diac/syncAnalysis}, provides a rigorous framework for analyzing both how BIPs unfold over time (diachronic dimension) and what constitutes the experience of a BIP at a given moment (synchronic dimension), as well as comparing them across participants, which fits with the goals of our study.

\section{Study: exploring breaks in presence through micro-phenomenology}

Our overall experimental approach is to induce BIPs with distractors of PI or Psi, and use micro-phenomenology to investigate how they have been experienced by our participants. Since breaks in PI or Psi can occur spontaneously without being deliberately triggered~\cite{Pouke_2022_qualitative}, we also adopt a bottom-up approach to data collection and analysis, allowing participants to describe any BIPs that emerge, including those we do not intentionally induce.

\subsection{Inducing breaks in presence}
\label{sec:BIP}

\paragraph{The VR environment}

\begin{figure}[!] 
\centering
 \includegraphics[width=\linewidth]{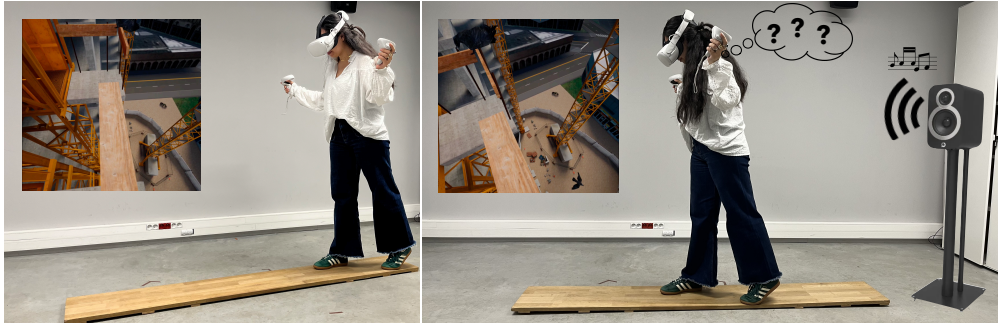}
 \caption{We immersed participants in a VR environment designed to induce a strong sense of presence by leveraging the fear of heights. The setup included a virtual plank, accompanied by a real plank calibrated to match the virtual one so as to reinforce the sense of presence. We introduced events likely to cause breaks in presence, such as loud music.}
 \label{fig:VRplank}
\end{figure}

We required a VR environment capable of inducing a strong sense of presence, where we could trigger BIPs. It had to be simple (not relying on controller-based locomotion), engaging, non-violent, standalone (to avoid cables), and free. We found in the literature that subjective feelings of fear in a VR environment can be positively correlated with a sense of presence~\cite{wilkinson_minireview_2021, maymon_presencefear_2024}. We selected an environment originally created for fear of heights exposure therapy, called \textit{EVEREST}, that was used in a previous experimentation~\cite{bulteau_feasability_2022}.

\begin{figure*}[t!] 
 \centering
 \begin{subfigure}{0.32\textwidth}
 \includegraphics[width=\linewidth]{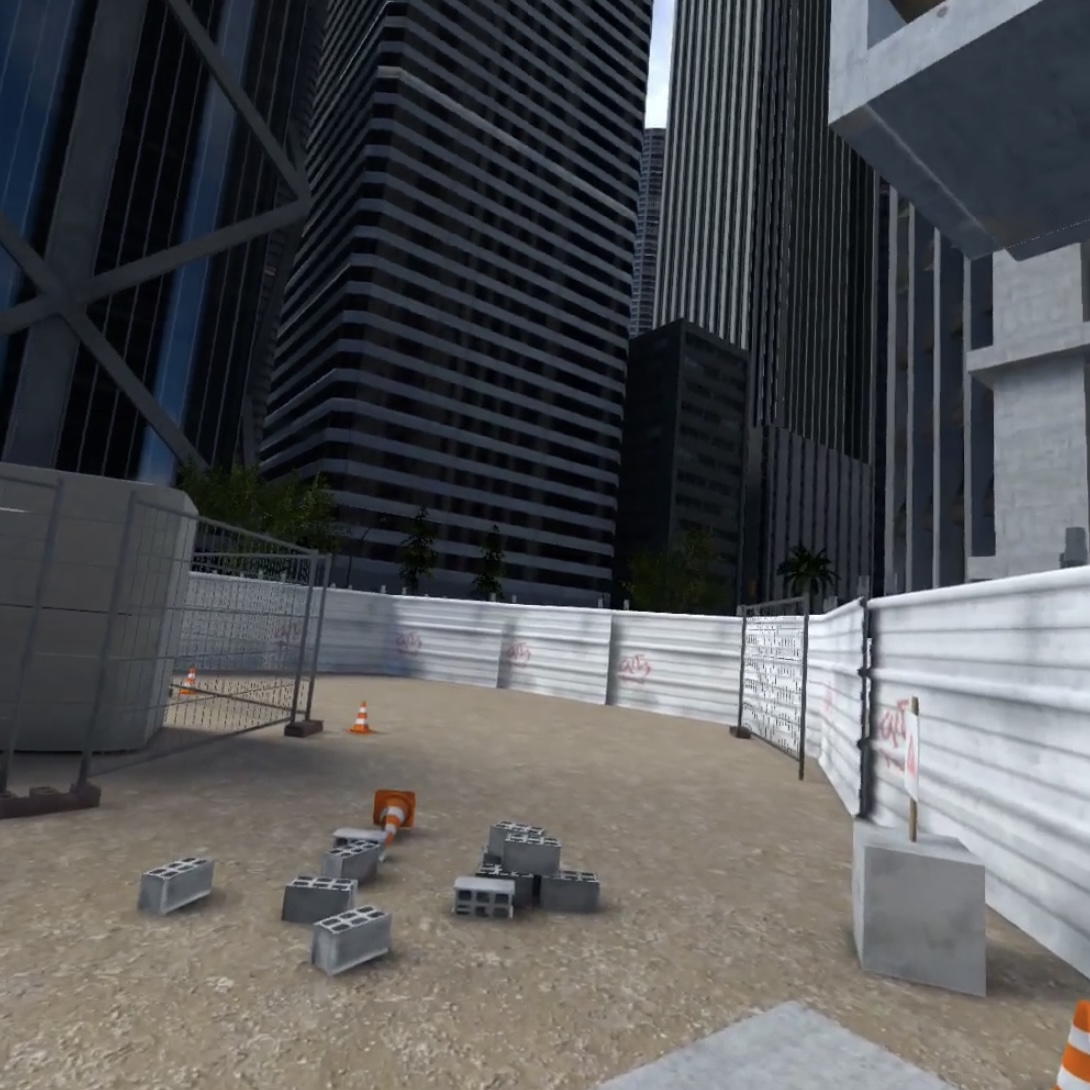}
 \caption{}
 \label{fig:start}
 \end{subfigure}
\hfill 
 \begin{subfigure}{0.32\textwidth}
 \centering
 \includegraphics[width=\linewidth]{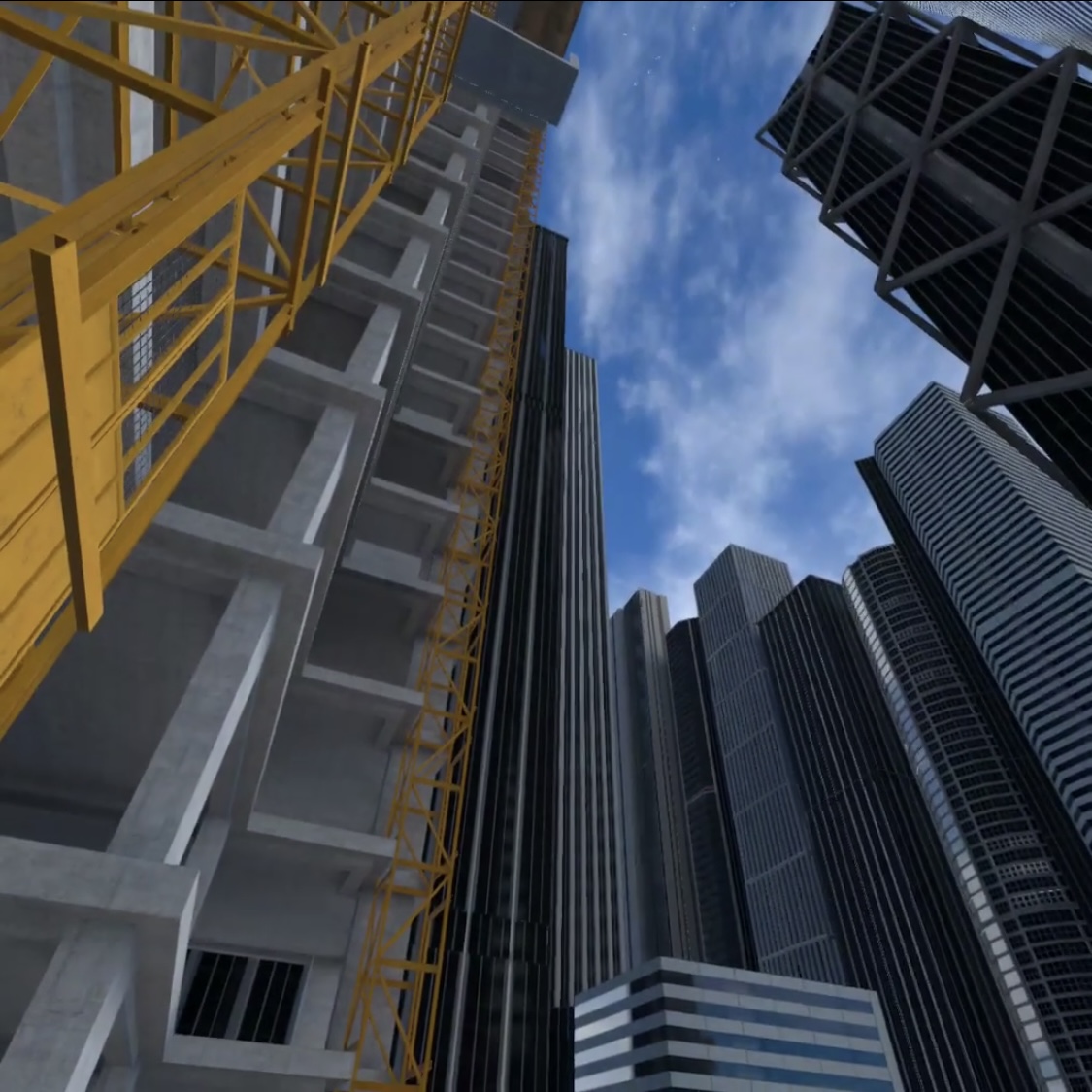}
 \caption{}
 \label{fig:skycrapper}
 \end{subfigure}
\hfill 
 \begin{subfigure}{0.32\textwidth}
 \includegraphics[width=\linewidth]{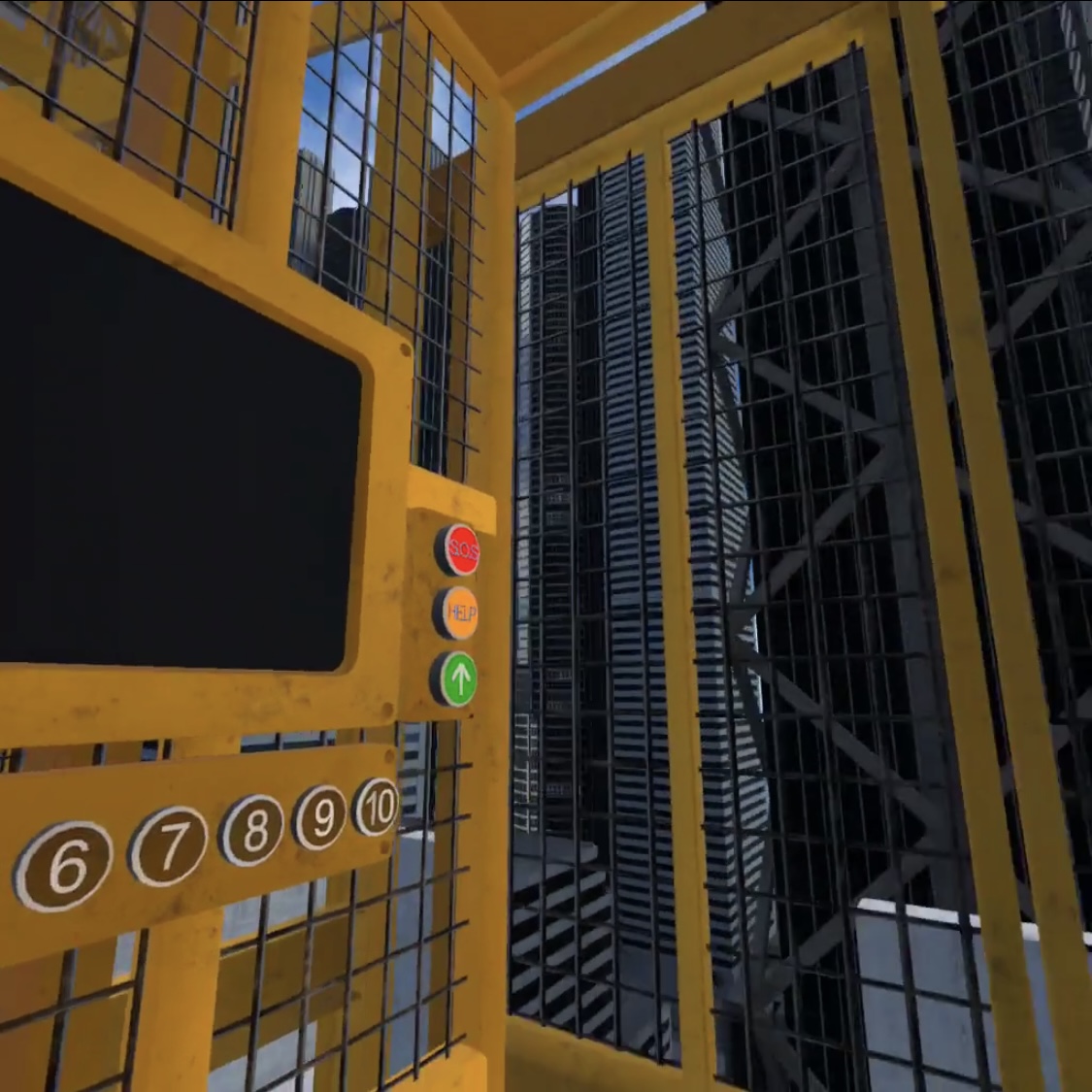}
 \caption{}
 \label{fig:elevator}
 \end{subfigure}
 \caption{\textit{EVEREST}: a VR environment originally created for acrophobia exposure therapy. (a) The starting point of the environment. (b) The cityscape around the participant in VR. (c) One of the two lifts used to go up. The green button starts the elevator, the orange button calls the experimenter and the red button takes the participant to a ``safe place''}
 \label{fig:image2}
\end{figure*}

\textit{EVEREST} consists of an environment featuring a 150m skyscraper under construction, with two elevators opening on external platforms. It includes a realistic cityscape surrounding the construction site (Fig.~\ref{fig:start}), emphasizing the building's height (Fig.~\ref{fig:skycrapper}). Users can take one of the elevators (Fig.~\ref{fig:elevator}), which brings them to a small platform, from which they can cross a plank to reach a second platform and a second elevator~(Fig.~\ref{fig:VRplank}). They are instructed to move up the building, by crossing the plank at each stop of one elevator. The environment can be parameterized, which allows to change the appearance and the size of the platform, the appearance of the plank, to add objects, or to trigger various events at different floors. 
Participants can stop the session at any time by pressing a red button in the elevator or several buttons on the controllers. To reinforce the sense of presence, a real wooden plank was placed in the experimental room, and calibrated to match the virtual plank traversed in VR (Fig.~\ref{fig:VRplank}).

\paragraph{Provoking BIP} 
\label{sec:provokingBIP}

As stated earlier, we drew on two models to design our distractors. The first is the PI/Psi model~\cite{slater_place_2009}, which provides actionable ways of breaking presence. 
Disrupting PI is relatively straightforward, as it involved disrupting immersion. Disrupting Psi is more difficult: because we could not modify the overall coherence of EVEREST (which would have required extensive development), we tried to introduce  inconsistencies in the virtual world.
The second model we relied on is that of~\citeauthor{chertoff_improving_2008} (see Section~\ref{sec:interalExternalBIP}), which identifies five concrete causes of BIPs~\cite{chertoff_improving_2008}:

\begin{itemize}
 \item External distractors, particularly external sounds, appear to be the most prominent sources of BIPs reported in the literature~\cite{slater_virtual_2000, liebold_continuous_2017, Savalle2024}. We designed various events that relied on the auditory channel to distract participants from the virtual environment and shift their cognitive resources back to the external world, since the sound originated outside the VR environment. We triggered an external sound at step 5 (10th floor), where we played four seconds of ``I Will Survive'' by Gloria Gaynor in the experimentation room. At step 14, we called the participant's name, a personalized strategy known to effectively capture auditory attention~\cite{holtze_name_2021}. 
 
 \item Inconsistent mediation BIPs arise when the environment fails to consistently support its intended message or media. They typically appear when a program freezes or a graphical bug arises, like a black screen. At step 16, we induced three brief black screens (each lasting 0.5 seconds) after the participant crossed the plank\footnote{This distractor exploited a bug in the HMD monitoring application, which was fixed by META two weeks into the study, preventing its use with participants P7 to P14.}. Similar to music, this distractor was intended to break the immersiveness of the environment and thereby induce a break in PI.

 \item Contradictory mediation BIPs occur when the mediated environment contradicts participants’ established schemas. At step 8, we removed the physical plank that participants relied on to cross the virtual plank, creating an unexpected absence, as well as potentially increasing anxiety, as the missing plank could be perceived as a loss of safety or support during the crossing. This distractor was designed to disrupt PI, since we expected participants to look for the physical plank and realize that it had been removed in real life, but also Psi, as the participants expected to feel a plank under their feet, but its absence created an incoherence with the constructed plausibility of the VR world.

 \item Unrefined mediation BIPs occur when the mediated environment triggers an excessive number of competing schemas. At step 17, after the elevator ascended multiple floors, we introduced helicopter noise, flying birds, and a floating tarpaulin in an attempt to provoke a cognitive overload. This distractor was designed to emphasize the ``surreal'' nature of the environment, thereby disrupting the coherence of the virtual world and, consequently, breaking Psi.

 \item Internal distractors arise when the user’s internal world interferes. They are inherently more difficult to induce and control, as they depend on the participant’s psychological state. We did not induce such breaks but remained attentive to BIP that could have them as causes during data collection.

\end{itemize}

~\\

Table~\ref{tab:BIP+sensoryAnchors} summarizes our distractors. We ensured that they were separated by plank-crossings and/or long elevator periods (for the last distractor), so as to give participants time to re-establish their sense of presence. We chose not to randomize the distractors, because in this exploratory study we focused on the precise description of each BIP individually, and were not interested in the way those may evolve during the entire VR session. Such experimental design was successfully employed in prior research on BIPs~\cite{liebold_continuous_2017}. Before conducting the main experimentation, we performed six pilots to test and fine-tuned the distractors, refine the experimental procedure, and improve the interviewing procedure.

\begin{table}[ht]
\centering
\begin{adjustbox}{width=1\textwidth}
\small
\begin{tabular}{|c|c|l|l|}
 \hline
 \textbf{Step} & \textbf{Floor} & \textbf{Distractor} & \textbf{Participant position when distractor is triggered}\\
 \hline
 5 & 10& Music & Crossing the plank \\
 \hline
 8 & 13& Plank removed & Leaving the first elevator \\
 \hline
 14 & 36& Knocking + Name of the participant & Entering the second elevator \\
 \hline
 16 & 43& 3 short black screens & Entering the second elevator \\
 \hline
 17 & 50 & Helicopter + Seagulls + Tarpaulin & Leaving the first elevator \\
 \hline
\end{tabular}
\end{adjustbox}
\caption{Distractors used during the experimentation to induce BIPs}
\label{tab:BIP+sensoryAnchors}
\end{table}

\begin{figure}[h]
 \centering
 \includegraphics[width=0.4\linewidth]{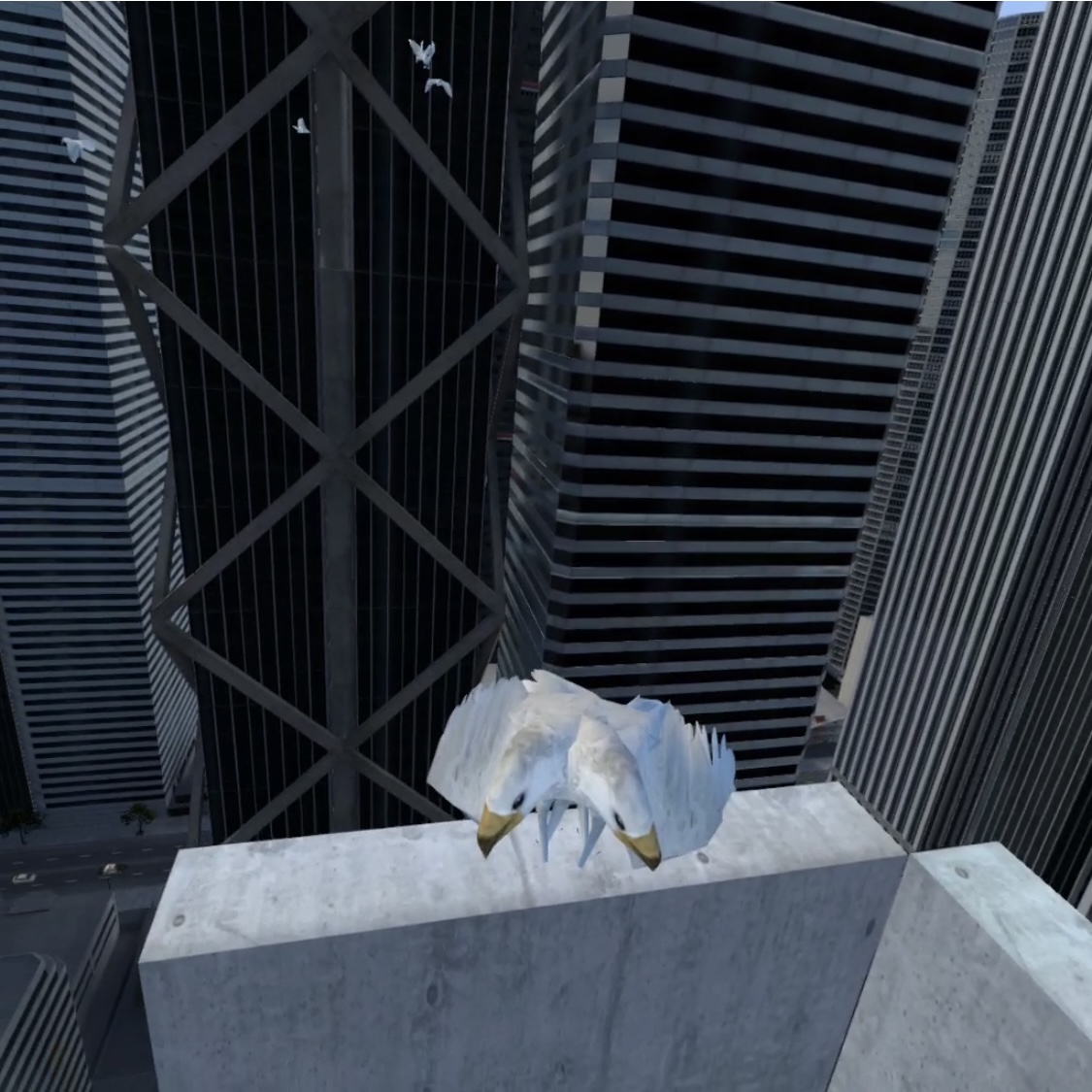}
 \caption{A two-headed seagull in the environment that caused unexpected BIPs}
 \label{fig:seagull}
\end{figure}

\subsection{Participants and recruitment}

We recruited 14 participants (11 women, 2 men, and 1 non-binary), aged from 20 to 56 yo $(mean~=~34.86, SD~=~11.51)$, through social media and flyers distribution at various locations. We did not compensate them.

The environment could induce a strong emotional response for those presenting a fear of heights, and we asked candidates to answer two standardized questionnaires: the Visual Height Intolerance Severity Scale (vHISS)~\cite{huppert_questionnaire_2017} and the Acrophobia Questionnaire (AQ)~\cite{abelson_cardiac_1989}. We only selected participants who had a score of less than 45 on the AQ and less than 7 on the vHISS (based on \cite{bulteau_feasability_2022}). The results of these questionnaires can be found in appendix~\ref{app:vHISSandAQ}. The protocol was validated by our local IRB (IORG0011023) under reference 22052024-1.

\subsection{Apparatus}
\textit{EVEREST} ran on an autonomous Oculus Quest 2. We recorded the interviews and the first-person VR video using a 13" MacBook Pro M1 running Mac OS Sonoma (14.5).

\subsection{Procedure}

Each session lasted about one hour, including the interview. Participants who completed the VR task progressed through 19 steps, spanning from floor 1 to 54\footnote{Between certain steps, the elevator ascended multiple floors.}, with 5 specific distractors designed to induce BIP (Table~\ref{tab:BIP+sensoryAnchors}).

\subsubsection{Welcoming participants}

The participants were taken to a soundproof test room, which had been prepared in advance (calibration of the physical plank). They signed a consent form and completed a demographic questionnaire, which included questions about their use of VR, meditation habits, and sports practice. We presented them with the VR environment and the task, without mentioning that we were looking for BIPs. We informed them about the ``safe place'', and that they could stop the experimentation at any time. 

\subsubsection{VR session with induced BIPs}
\label{sec:descriptionOfBIP}
We set up the VR headset and controllers, and guided the participants through the VR environment for the first few seconds. Then, we informed them that their questions could no longer be answered. During the VR session, the experimenter intentionally provoked the first four distractors (see section~\ref{sec:BIP}).

\subsubsection{Presence questionnaire}

Immediately after completing the VR part of the experimentation, we administered a presence questionnaire to assess whether the virtual environment effectively elicited a high level of presence~\cite{xiao_concept_2025}. We choose the Presence for Immersive Environments (SP-IE) questionnaire~\cite{khenak_construction_2019} (without the optional dimension on social presence with avatars), for several reasons: it is validated in French, it is structured around six dimensions relevant to our study (see \ref{sec:QuestAnalysis}), and participant have less difficulty answering it in comparison to the Presence Questionnaire~\cite{witmer_measuring_1998}\footnote{During the pilot studies, we administered both the SP-IE and PQ questionnaires, and participants reported difficulties understanding several items of the PQ, particularly questions 18 and 19. This was likely due to PQ having been designed with an industrial engineering perspective, which may not align with our task, where the only performance indicator is being able to cross the plank.}.

\subsubsection{Interviews}
\label{sec:interview}

\begin{figure}[h]
\centering
 \includegraphics[width=0.8\linewidth]{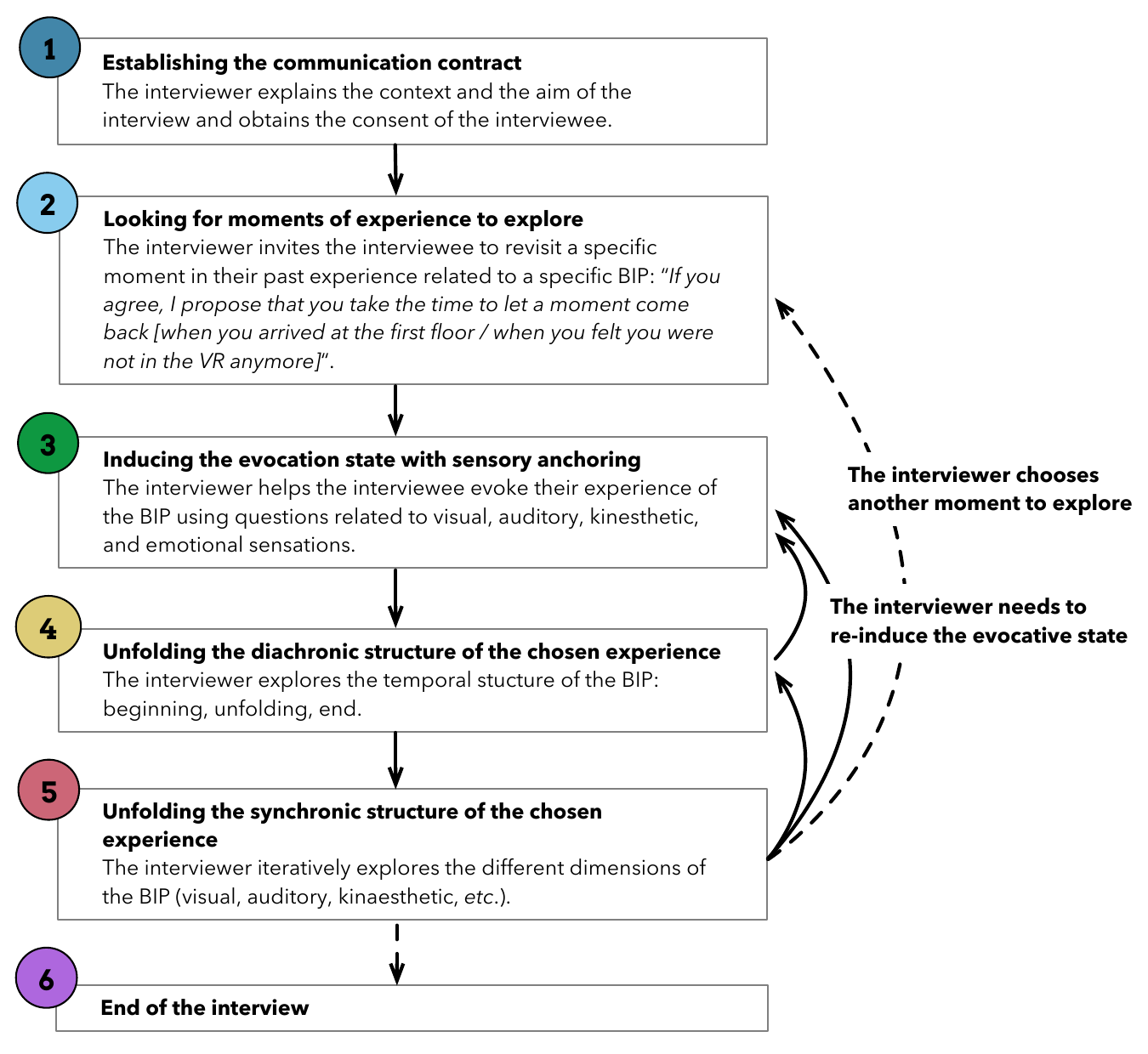}
 \caption{The phases of a micro-phenomenological interview follow a sequential order but may require iterations between phases 3, 4, and 5 (solid arrow). Once a moment of interest has been sufficiently described, the interviewer may either choose to end the interview or explore another moment of interest (dashed arrow). Figure inspired by~\cite{hogan_elicitation_2016}.}
 \label{fig:MP}
\end{figure}

\subsection{Data Collection}

We recorded the participant point of view for each VR session. We also video-recorded the interviews. We assigned each participant a unique ID. In what follows, we refer to each participant with a number, e.g., P5 corresponds to participant~5.

\subsection{Data Analysis}
\label{sec:dataAnalysis}

13 out of 14 participants successfully completed the task and reached the floor corresponding to the 19th step. P5 chose to quit VR after four steps (i.e., before the music distractor) due to an intense fear of heights. We chose to include her in the study since she had nevertheless experienced two BIPs, had been interviewed on these, and did not object to her data being retained. 

\subsubsection{Questionnaires analysis}
\label{sec:QuestAnalysis}
The 18 questions of the SP-IE questionnaire were analyzed according to the 6 dimensions it measures, as defined in \cite{khenak_construction_2019}. For each dimension, we summed the scores of the relevant questions\footnote{Sense of spatial presence (Q.3, 5, 7, 9), affordance of the environment (Q.11, 12, 14, 16), enjoyment (Q.1, 2, 6), sense of reality of the environment (Q.10, 15, and 17), attention allocation to the task (Q.8, 13), and cybersickness (Q.4, 18).} and calculated averages and standard deviations.

As expected, results showed an overall good level of presence (considering that 4.5 is a very high level of presence~\cite{khenak_spatial_2020}), particularly for sense of spatial presence $(Mean=3,79; SD=0,73)$, affordances of the environment $(Mean=4,27; SD=0,44)$ and attention to the task $(Mean=4,27; SD=0,70)$. Enjoyment $(Mean=3,46; SD=0,63)$ and realness $(Mean=3,49;SD=0,66)$ were a little lower, with some finding the unsecured under-construction setting unrealistic or unpleasant. Despite this, most participants remained fully engaged in the task (except P5), and reported low cybersickness $(Mean=1,81; SD=0,69)$. Appendix \ref{app:secA1} details these results.

\subsubsection{Interview transcript analysis}
\label{sec:diac/syncAnalysis}

We analyzed our interviews following a diachronic/synchronic
analysis procedure inspired by~\cite{valenzuela_procedure_2019}. This bottom-up method aims to model the collected experiences as diachronic and synchronic structures, that can be further compared and
abstracted. We chose it for three reasons: it ensures that the diachronic and the synchronic richness of the collected experiences is not lost, it documents the criteria used
for building the descriptions, and it enables comparison between participants. The method includes 5 steps, to which we added normalization steps:

\textbf{Step 1: Data preparation.} We transcribed the recordings and used the video of the interview to describe gestures or gazes of the interviewees. Then, we discarded the verbatims that we did not consider as descriptions of authentic experience~\cite{petitmengin_validity_2009}. For this, we used what was said (e.g., present tense is a cue of authenticity), the questions of the interviewer (e.g., ensuring they were not inducing the answer), and the gestures/eye-directions (e.g., describing action with the hands is a cue of authenticity, as is looking up and sideways while being in contact with the past). 

\begin{figure*}[ht!] 
 \hspace{-3cm}
 \includegraphics[width=1.4\linewidth]{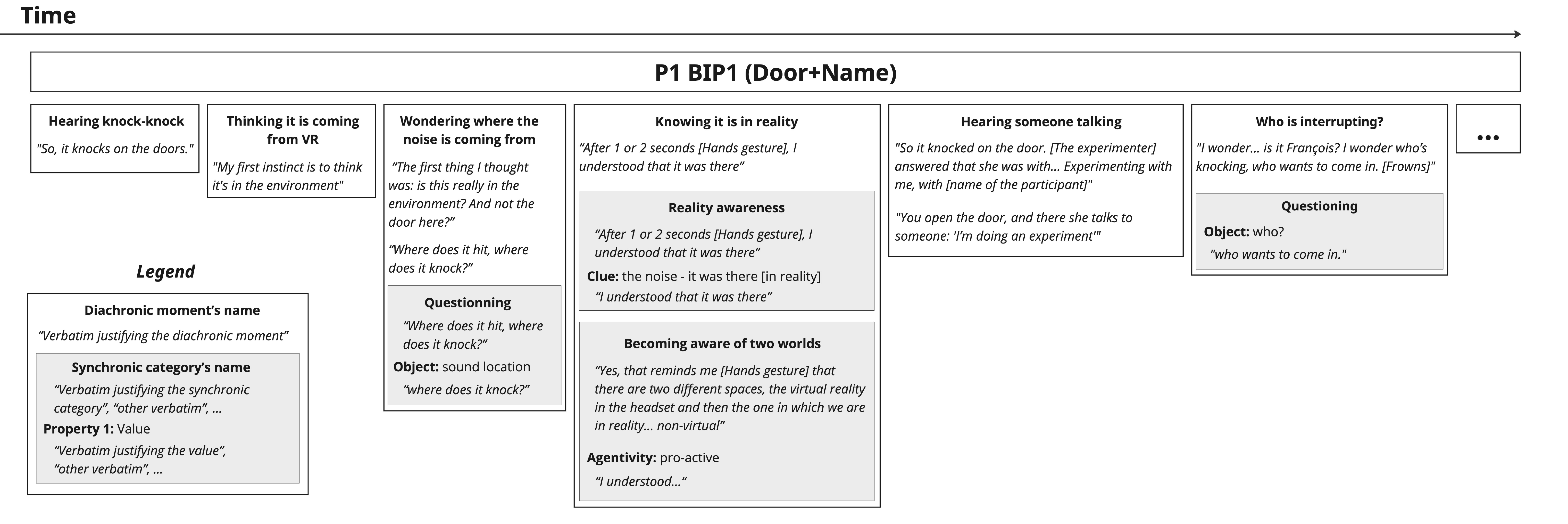}
 
 \caption{
 Example of a diachronic/synchronic model resulting from the analysis (step 1 to 3) of the experience of P1 with the $\mu$PMT  software. Verbatims from the interviews are classified into diachronic moments (white boxes), which are organized sequentially (diachronic analysis). Afterwards, moments are further described with synchronic categories represented in grey, possibly with properties, also justified by verbatim (synchronic analysis). The diachronic and synchronic structure of each modeled BIP is called a BIP episode. }

 \label{fig:upmt}
\end{figure*}

\textbf{Step 2: Specific diachronic analysis.} We analyzed the temporal evolution of each described lived experience and constructed a temporal structure for each potential BIP by: 1) grouping together the verbatims related to the same diachronic moment, and 2) organizing these moments chronologically. When necessary, we relied on first-person videos to support this process, ending with one diachronic structure for each BIP that had been described by each participant (e.g., \textit{P1 BIP1 (Door + name)} on Fig.~\ref{fig:upmt}).

\begin{figure}[h]
 \hspace{-2.5cm}
 \includegraphics[width=1.5\linewidth]{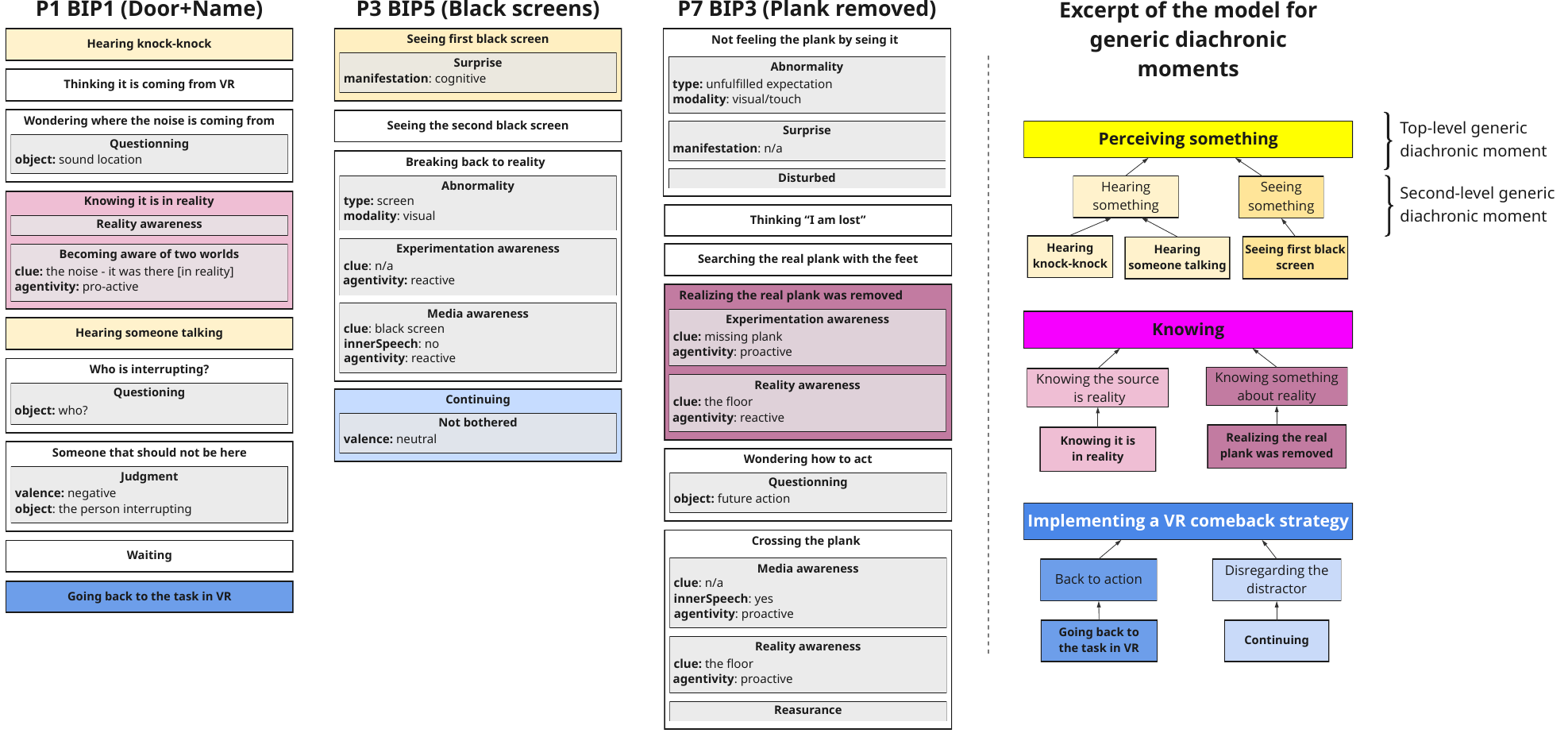}
 \caption{
 Left of the dotted line: the unfolding of three BIP episodes, from top to bottom (note P1BIP1 is also described in Figure~\ref{fig:upmt}). Right: excerpts of the model normalizing some diachronic moments that can be found in these episodes. 
 During step 4 of the analysis we grouped the BIP moments' names into more abstract categories, using color coding: second-level generic diachronic moments have a gradient variations of their top-level generic moments.}
 \label{fig:analysisMethod}
\end{figure}

\textbf{Step 3: Specific synchronic analysis.} When the information was available in the verbatims, we further described the moments using synchronic categories corresponding to what the participant had mentioned experiencing (e.g., the synchronic category \textit{Reality awareness} for P1 BIP1 has a property \textit{Clue}, Fig.~\ref{fig:upmt}). Fig.~\ref{fig:upmt} illustrates part of the diachronic/synchronic structure created after step 3 of the analysis, that we called a \textit{BIP episode}. 
Fig.~\ref{fig:analysisMethod} describes three other examples of BIP episodes (aligned vertically) with their diachronic moments and associated synchronic categories.

At the end of Step 3, we concluded that some of the modeled episodes could not be considered genuine BIPs because they did not involve a break in either PI or Psi. We therefore excluded these episodes.  
We considered an episode to be a genuine BIP when participants reported one of the following: not being fully ``there'' in the virtual environment (break in PI); an intrusion of the real world into their lived experience (break in PI); perceiving something as implausible or incoherent with the VR world (break in Psi); or describing a situation similar to all the others already reported.

\textbf{Steps 4: Generic diachronic analysis.} We regrouped the names of all the kept moments from all the episodes thematically, to create a hierarchy of more and more general diachronic moments. For example, when P1 reported ``Hearing knock-knock'' and later ``Hearing someone talking', we grouped these two instances into a (second-level) generic diachronic moment called \textit{Hearing something}, which in turn fell under a top level diachronic moment called \textit{Perceiving something}. Fig.~\ref{fig:analysisMethod} presents an example with three episodes and the associated model excerpts. Fig.~\ref{fig:topLevelMoments} presents the resulting hierarchy.
The color codes were used in a global visualization of all BIP episodes to help identify and group similar BIP patterns (Fig.~\ref{fig:BIPGlobal}).

\textbf{Steps 5: Generic Synchronic analysis.} In the last step, for each BIP pattern we had identified, we considered the most used synchronic categories (see Table \ref{tab:synchronic}) and analyzed how they were used across BIP patterns. This enabled us to refine the analysis and better describe each pattern.

\paragraph{Normalization.} Throughout the whole analysis process, we conducted multiple sessions to check, discuss and settle on the specific modeling of each BIP (organization of the moments, names, etc.); and to normalize the generic categories we would use in our generic analysis. These alignment sessions were essential to use a consistent terminology across different participants and BIPs, allowing to compare the diachronic and synchronic structures of BIP episodes (steps 4 and 5). 
We also standardized the naming convention for each BIP: P\# BIP\# ([Name of the BIP]), where P\# represents the participant ID and BIP\# indicates the order of the BIP within the session. The name of each BIP (induced or not) can be found in Table~\ref{tab:BIP}.

\paragraph{Iterative process}
We followed an iterative analysis process where we started the data analysis in parallel of data collection, following the recommendations of~\citeauthor{petitmengin_discovering_2019}~\cite{petitmengin_discovering_2019}\footnote{Section 5.2 ``\textit{The mutual guidance of structural and experiential unfolding processes}''.}. We conducted three iterations of analysis, during which we completed steps 1 through 3 each time: the first iteration was carried out on the first 4 participants, then we iterated on the next 6 participants, and finally on the last 4 participants. This allowed us to refine our analysis twice and to guide subsequent data collection. For example, after the second interview, we noted that a double-headed seagull actually caused a BIP (see Fig~\ref{fig:seagull}). This event was then investigated in subsequent interviews as a possible BIP. 
We completed the generic diachronic and synchronic analysis (steps 4 and 5) after all the data had been collected. In some cases, discussions on generic analysis (e.g., merging two synchronic categories) led us to get back to specific analysis and modify BIP episodes, so as to ensure coherence at all the levels of analysis.

\paragraph{Data coders}
For practical reasons, the first author conducted 13 interviews and the second author conducted one interview\footnote{Both interviewers were trained in the technique.}.
Steps 1, 2, and 3 were performed by the third author only, but discussed with all the members of the team during next steps. The normalization sessions as well as the generic diachronic and synchronic analysis (steps 4 and 5) were conducted by all the authors to ensure consistency and strengthen the validity of the findings. During these two steps, we frequently referred back to verbatims that described our modeling elements, to the raw transcriptions, and to the BIP episodes, so as to ensure the validity of our decisions.

\paragraph{Tools for the analysis}

Each interview analysis was carried out using the $\mu$PMT tool\footnote{ Micro-Phenomenology Modeling Tool: \url{https://yprie.github.io/uPMT/}.}, which allowed us to manage the interviews, highlight verbatim, organize sequences of moments, and manage a list of synchronic categories and diachronic moments. Although $\mu$PMT did not automate the analysis processes, it facilitated the creation of models by allowing easy reorganization of verbatims, moments, or properties with drag and drop. 
For steps and 4 and 5, we used a classical spreadsheet, always keeping the coherence between the spreadsheet and the specific models in $\mu$PMT.

\section{Results}

We modeled 57 BIP episodes, each representing the lived experience of a BIP by a participant\footnote{On average, each participant experienced 4.15 BIPs $(SD~=~1.63, min~=~1, max~=~6)$.}. Each BIP episode includes at least 3 diachronic moments (max. 19) with an average of 7.23 diachronic moments per BIP episode $(SD~=~3.61)$. From these BIP episodes, we created 8 top-level and 23 second-level generic diachronic moments, which form the building blocks of our generic diachronic analysis of BIPs episodes, through which we identified 4 distinct patterns. 

In this section we illustrate some of the BIPs we found, before detailing the 8 top-level generic diachronic moments we found, the most frequently used synchronic categories, and 4 patterns of BIPs episodes.

\subsection{Modeling breaks in presence}

Regarding the origins of the BIPs, while the majority were triggered by our distractors (37/57), more than a third (20/57) arose spontaneously from the participants (see Table~\ref{tab:BIP}). For instance, P6 described a mismatch between visual and sensory cues, explaining that seeing the tarpaulin move atop the building without feeling any wind created a sense of inconsistency (indicating a possible break in the Plausibility Illusion). Let us describe two examples of BIPs, before explaining why we classified them as breaks.

A first example of BIP is related to what happened to P1 when he heard a knock on the door. At first, he believed the sound originated from the virtual environment, but quickly began to question whether it might have come from the real world. As P1 realized it was someone knocking on the door of the room, he described the coexistence of two distinct worlds: ``\textit{It reminds me [hand gesture] that there are two different spaces: the virtual reality and the one we're in, reality... The non-virtual reality.}''. When he heard the experimenter speaking, he wondered who was interrupting. He chose to wait until the interruption ended, and then came back to his task.

Another example of BIP occurred when we removed the physical plank for P7. She reported stepping onto the virtual plank when, unexpectedly, she could no longer feel the real one under her foot. This sensory mismatch surprised and unsettled her. Her initial reaction was disorientation and she believed she had lost her position in the room. She began searching for the real plank with her feet but quickly realized that, if the virtual plank was visible, the physical one was missing. She explained: ``\textit{I put my foot forward, but I can't find anything [...] But visually, it's here!}''.
She then deduced that the real plank had been intentionally removed: ``\textit{Blast, they took it away and I didn't hear it. [...] It pulled me out; it forced me to remember that I was in reality}''. She questioned herself on how to proceed, and started to reassure herself: ``\textit{It's harder to move forward because I can't feel anything under my feet. [...] I tell myself that, I saw earlier, that the ground is solid. [...] I know it's there. [...] I tell myself, you can walk, you won't fall}''.
Ultimately, she managed to cross the (virtual) plank, regaining enough confidence to complete the task.

These two examples clearly describe BIPs. P1 explicitly acknowledged the real world, describing himself as being ``\textit{between two worlds}''. This indicates a break in Place Illusion, as the participant no longer experiences a full sense of being located within the virtual environment. In the second example (P7), the participant reassured herself that the situation was not real in order to cross the virtual plank. This episode contains clues of both PI and Psi. Indeed, the sensory mismatch points toward a break in Psi (``\textit{I put my
foot forward, but I can’t find anything [...] But visually, it’s here! }''), while the quote ``\textit{it forced me to remember that I was in reality}'' is more indicative of a break in PI.

\input{tables/tbl_BIP}

\begin{figure}[H]
 \includegraphics[width=1\linewidth]{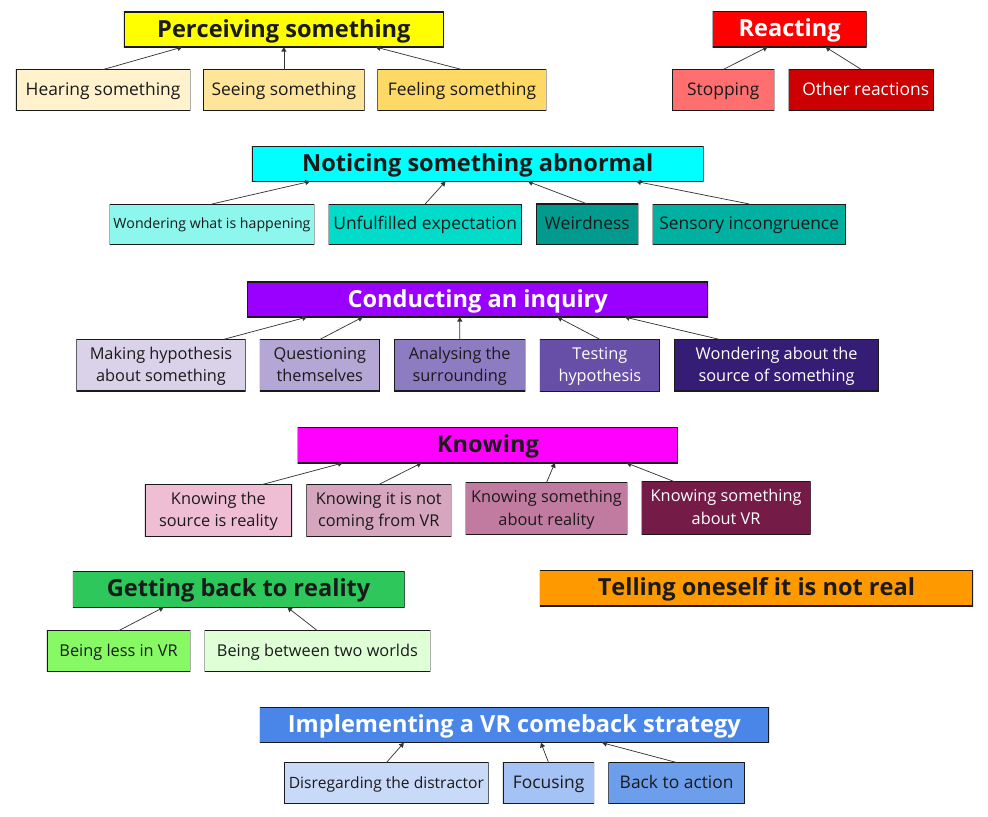}
 \caption{Top-level generic diachronic moments and their associated second-level diachronic moments (with gradient color variations of the top-level color).}

 \label{fig:topLevelMoments}
\end{figure}

\subsection{Generic diachronic moments}

From all the modeled BIP episodes, we identified eight top-level generic diachronic moments (see Fig.~\ref{fig:topLevelMoments}).
Here we detail each of them, along with its associated second-level generic diachronic moments.

\input{tables/tbl_BIPRU}

\paragraph{Perceiving something} 
\label{sec:perceiving_something}
This top-level generic diachronic moment describes the first perception of a distractor by a participant. It corresponds to the moments when participants hear, see, or feel something. For example, P4 explains hearing music: ``\textit{At one point there was music [...] A ringing cell phone, something like that}'' (P4). A large part of BIP episodes (40 occurrences) started by the perception of a distractor. 
Interestingly, in some cases, distractors were perceived by participants but interpreted as being part of the VR environment. For example, P14 explained hearing music, but not living it as a BIP: ``\textit{I finished crossing the plank and I was walking to the elevator. [...] When I hear it... It was music... Someone singing, there was a voice. It was close, though, [...] and I thought maybe it was the sound of a building across the street...}'' We did not consider such experiences as BIP episodes.

\paragraph{Noticing something abnormal}
Participants noticed that something abnormal or incongruous was happening in 23 BIPs. In 14 instances, they experienced \textit{unfulfilled expectations}, where something they anticipated did not occur. For example, P3 noted: ``\textit{The sensation is not the same compared to the other floors where I felt the plank under my foot. [...] The fact of having a different floor [...] Here, I didn't feel anything}''. In 6 cases, participants encountered \textit{sensory incongruence}, where two sensory perceptions conflicted. P2 described this experience: ``\textit{I couldn't see anything, [...] but I heard [hand gestures] that there was something there}''. In another 6 cases, participants reported a sense of \textit{weirdness}, as illustrated by P6 observing the two-headed seagulls: ``\textit{The bird, I see it has the body of one, but that it has two heads}''. All these elements point toward a break in the Plausibility Illusion.
The moments subsumed under the top-level generic category \textit{Noticing something abnormal} are mainly characterized by instances of the synchronic categories abnormality (12 occurrences), surprise (5), and weirdness (2). 

\paragraph{Reacting}
Moments subsumed by this generic synchronic category correspond to physical reactions to distractors. They can be found for 9 participants in 14 BIPs. Most of these reactions consisted in \textit{stopping any action} (9 occurrences): ``\textit{I had to wait a while to find out if it was something that influenced what I did or not}'' (P13). \textit{Other reactions} were classified as involuntary reactions and included laughing (P12, P13), being thrown off balance (P6, P12), and turning the head (P2, P9). 
Reacting moments such as stops were difficult to precisely locate, as participants' generally could not pinpoint the exact moment it ended. Although they likely extend across multiple diachronic moments, we categorized these as events that facilitate pattern matching.

\paragraph{Conducting an inquiry}
A key aspect of many BIPs (27/57) was that participants actively sought to make sense of what was happening. In many cases, this involved a process of inquiry. Various strategies were employed, the most common (15 occurrences) being \textit{making hypotheses} about the distractor. For example, P1 hypothesized about the black screens, explaining: ``\textit{I said to myself, 'Damn, maybe there's a battery problem. [...] Was the headset properly charged before the experimentation? Will it shut down?'}''. Some participants \textit{analyzed the surroundings} (4) or \textit{wondered about the source} of the distractor (5), as P13 described: ``\textit{At first, I wasn't sure... since the knocking sound was brief, I couldn't immediately tell whether it was coming from the real-world or the virtual reality environment}''. Other participants were \textit{questioning themselves} on what was happening (8) or \textit{tested a hypothesis} (3). For example, P13 searched for the missing plank, explaining: ``\textit{I look a little further, thinking maybe I am in the wrong spot. I check a bit more to make sure it isn't just me who makes a mistake. I check with my foot}''.

\paragraph{Knowing}
We found that participants were able to identify the distractor and understand why it occurred (27 occurrences). In some of these instances (14/27 occurrences), this realization came as a result of participants' inquiries into the source of the distractors. For example, P13 explains how her inquiry ended on the conclusion that \textit{the noise was not coming from VR}: ``\textit{She spoke [the experimenter]; I knew it wasn't in virtual reality then}''. In other cases, participants did not engage in any inquiry, immediately recognizing that \textit{the source is from reality}. P11, for instance, quickly identified music as a phone ringing in the experimentation room: ``\textit{And then, as it was ringing, I said to myself that it was a telephone}''. Some participants relied on their knowledge of reality (or VR) to precisely identify the distractor and deduce what was happening. For example, P1's reaction after hearing a knock on the door: ``\textit{Here's someone who wants to get into the room when it's not the right time...}''.

\paragraph{Implementing a VR comeback strategy}
Participants employed strategies to re-immerse themselves in VR (37/57 BIPs). We identified three distinct strategies. The most common strategy (30 occurrences) was to \textit{get back into action}, by walking or pressing the elevator button. P4 explained: ``\textit{I want to go up one floor, so I need to continue and push the button; I get back into action}''. Another approach was to \textit{focus on the environment} by looking around, as P6 described: ``\textit{I try to stay in the virtual reality mood [...] I am looking around me}''. Another strategy involved deliberately disregarding the distractor, as P8 did: ``\textit{I said to myself: 'Well, there is music [...] but you have to go for it [crossing the plank]!' }''. It is important to note that these strategies are not mutually exclusive, and can be used one after the other, such as for ``P7 BIP1 (Music)'' in Table~\ref{tab:BIPRU}.

Strategies for re-engaging with VR are primarily characterized by the synchronic categories of volition (10/14 occurrences) and attentional focusing (17/31). Volition refers to the conscious act of making a choice or decision, while attentional focusing involves deliberately directing one's attention to something. Volition implies action, whereas attentional focusing is more of a cognitive process. Both indicate that returning to VR is intentional, and involves cognitive processes and physical actions.

\paragraph{Telling oneself it is not real}
We observed that some participants chose to reassure themselves about the unreality of their experiences (10 occurrences). For instance, P3 calmed herself before crossing the plank by thinking, ``\textit{Visually, there's a plank, and since this is an experimentation, there must be a floor}''. Notably, P5 used this strategy twice to successfully cross the plank. We did not find second-level diachronic moments for this top-level generic moment, but it remains significant nonetheless.

\paragraph{Getting back to reality} Many participants described specifically returning to reality during the experimentation (17 occurrences), which constitutes a clear break in Place Illusion. They used various terms to express this shift, including ``\textit{rupture}'' (P1), ``\textit{break}'' (P4), ``\textit{parasite}'' (P4), ``\textit{coming out}'' (P4), ``\textit{disconnection}'' (P3, P12), ``\textit{cut-off}'' (P3, P9), ``\textit{separation from VR}'' (P14), ``\textit{come-back down}'' (P6), and ``\textit{short-circuiting the experience}'' (P1).

We identified two second-level moments, the first one where participants described \textit{feeling not immersed} in VR (14 occurrences). For example, P4 described this break as: ``\textit{I am a little frustrated or irritated [...] this is completely breaking me in my thing. [...] it comes to parasitize the experience I have of it...}''. The second second-level diachronic category is less common and corresponds to the experience of \textit{being between two worlds} (3 occurrences), which corresponds to experiencing VR and reality simultaneously. For example, P13 explained feeling her body in reality while her mind remained in VR, stating: ``\textit{What I see is totally virtual reality. [...] However, in terms of body sensations, I'm more in reality [...]. I have a duality between the two sensations}''.

\input{tables/synchronic}

\subsection{Generic synchronic categories}

After examining diachronic categories, we can now turn to synchronic categories. Table~\ref{tab:synchronic} presents the most used ones, their frequency and the number of participants associated with each. 
The most frequently reported categories were Reality awareness (46 occurrences, reported by all 14 participants), Experimentation awareness (45 occurrences, 13/14 participants), and Media awareness (29 occurrences, 12/14 participants) indicating a strong and widespread awareness of either the reality, the experimentation setup, or the VR headset during BIPs. 
Participants also frequently described where they intentionally directed their attention (Attentional focusing, 29 occurrences), suggesting that during BIPs they usually point their attention to specific stimuli.

Next, several adjectives were used by participants to characterize their BIP experiences. These included expressions of Abnormality (23 occurrences), Surprise (14 occurrences), Questioning (23 occurrences), Weirdness (10 occurrences), Being bothered (10 occurrences), and Discomfort (9 occurrences). Finally, there are synchronic categories related to the task itself: Attention (23 occurrences), Judgment (22 occurrences), Fear (15 occurrences), Volition (14 occurrences), and Reassurance (9). Note that we use the term ‘attentional focusing' to refer to moments when participants actively shifted their attention toward something (usually a distractor), and attention to describe what they were actually attending to.

\begin{figure}[h]
 \hspace{-3.2cm}
 \includegraphics[width=1.5\linewidth]{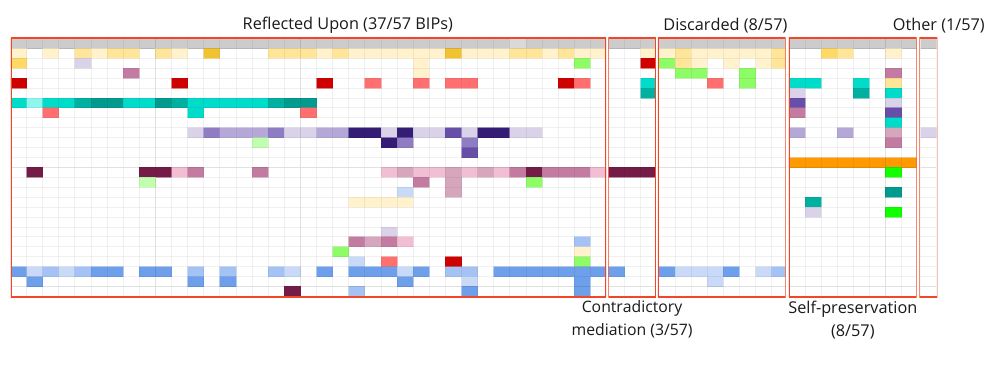}
 \caption{The 57 BIP episodes are grouped into 4 main BIP patterns. Each episode is composed of the diachronic moments sequentially arranged from top to bottom on one column (blank cells are only used for horizontal alignment). The color of each moment / cell corresponds to the associated second-level diachronic generic moment (see Fig.~\ref{fig:topLevelMoments}). Full table in supplementary material.} 
 \label{fig:BIPGlobal}
\end{figure}

\subsection{Generic diachronic patterns of BIP episodes}

By grouping specific diachronic moments into more generic ones, we were able to compare BIP episodes together. To do so, we organized all 57 BIP episodes vertically next to each other (see Fig.~\ref{fig:BIPGlobal}), and found four different main types: \textit{reflected-upon} BIPs (examples on table~\ref{tab:BIPRU}), \textit{discarded} BIPs (examples on Table~\ref{tab:BIPD}), \textit{self-preservation} BIPs (examples on Table~\ref{tab:BIPSD}), and \textit{contradictory mediation} BIPs. One BIP falls into the \textit{Other} category.

\subsubsection{Reflected-upon BIPs}
During such episodes, participants were consciously reflecting upon what was going on. 36/57 episodes correspond to this pattern, which is the most common in our data. Its particularity is to contain at least one of the top-level diachronic moments: \textit{Noticing something abnormal}, \textit{Conducting an inquiry}, or \textit{Knowing}. We found that these moments could occur several times in a row. For example, the first 2 BIPs of Table~\ref{tab:BIPRU} show that P1 and P12 heard a sound (knocking on the door; yellow), then conducted an inquiry (purple), which helped them identify the source of the distraction (pink). P12 attempted to ignore the disruption (blue). However, the distraction persisted, and both P1 and P12, continued to hear it. This prompted P1 to conduct another inquiry (asking who was interrupting) and to stop her movement. Ultimately, both participants applied one (or multiple) strategy to continue their task (\textit{focusing}, \textit{going back in action}, and \textit{disregarding the distractor}). In most cases (28/36), participants implemented a strategy to return to VR and re-immerse themselves.

In terms of synchronic categories, \textit{reflected-upon} BIPs typically exhibit a negative valence, leading participants to experience a loss of concentration (2 occurrences) and/or negative emotions such as being bothered (10), disturbance (8) or frustration (3), occurring at several diachronic moments. We also identified numerous occurrences of awareness-related synchronic categories: experimentation awareness, the awareness of being part of an experimentation (23); reality awareness, of the real-world (22); and media awareness, of VR technology (14).

\subsubsection{Discarded BIPs}
Such episodes occurred when participants did not experience any sense of abnormality, and dismissed the distractor without further investigation. These episodes only include the generic diachronic moments: \textit{Perceiving something} and \textit{Implementing a VR comeback strategy}. 8 episodes correspond to such a pattern, 4 of which are illustrated in Table~\ref{tab:BIPD}. For example, during BIP5, P3 reported seeing 2 black screens but rapidly disregarded this distractor: ``\textit{I didn't ask myself too many questions, I just kept going}''. Similarly, P4 and P8 reported hearing a knock on the door as well as their name being called. While both acknowledged that this disrupted their actions momentarily, they also reported they quickly dismissed the distractor and returned in VR.

With regard to synchronic categories, \textit{discarded} BIPs have a minimal impact on the VR sessions. Nevertheless, 4/9 episodes were associated with a synchronic category of negative valence: negative judgment (P3, P10), bothering (P4, P8), frustrating (P4, P8), and disturbing (P8). P4 describes her frustration as follows: ``\textit{It's extracting me from VR, [...] It's parasitizing me a lot. [...] It is frustrating...}''. In 8/9 episodes, participants also reported experimentation awareness.

\input{tables/tbl_BIPD}

\subsubsection{Self-preservation BIPs}
Many participants reported they had a fear of heights during the experimentation and consciously provoked \textit{self-preservation} BIPs, which correspond to the cognitive strategy of actively challenging the authenticity of one's experience. These episodes do not include a deliberate strategy to come back in VR as participants also experienced an involuntary return to presence due to their fear of heights. 

In terms of synchronic categories, this pattern is typically accompanied by strong indicators of presence, such as fear (9 occurrences), stress and anxiety (5). For example, P5 (P5 BIP1 in Table~\ref{tab:BIPSD}) initially perceived the plank as moving: ``\textit{I noticed that it was shifting a bit as I climbed onto it}''. Her immediate reaction was intense stress, reflecting a strong sense of presence: ``\textit{I am telling myself that I might pass out}''. To overcome this fear, she told herself that it was not real: ``\textit{I reassured myself by saying, 'No, don't worry, it's not real'} ''. Such self-reassurance, corresponds to the generic diachronic moment \textit{Telling oneself it is not real}. While these BIPs arose in response to fear, they were experienced positively by participants. Additionally, they were associated with media awareness (7 occurrences), experimentation awareness (6), and reality awareness (4).

\input{tables/tbl_BIPSD}

\subsubsection{Contradictory mediation BIPs}

We identified three BIPs where participants interacted or wanted to interact with the VR environment, typically to grab the guardrail in VR, but were unable to do so. The name \textit{contradictory mediation} was reused from~\cite{chertoff_improving_2008}. Such BIPs are characterized by the synchronic category of volition (e.g., the conscious intention to grab something) or by an involuntary reaction: ``\textit{I had this reflex to look for the wall on my left, the one I could see on the side}'' (P13). Additionally, participants reasoned (or found out in the case of P13) that the guardrail was absent, as explained by P6: ``\textit{I tell myself there's no point in trying to grab the wall, there isn't one}''. This caused them to experience reality awareness (3 occurrences) in all cases. Interestingly those BIPs carry a neutral valence (neither positive nor negative) and appear to hold little significance for the participants.

\subsubsection{Other BIP}
One episode did not fit in our patterns. P14 described experiencing a BIP in an elevator, saying, ``\textit{I wondered what would happen next, would there be something else?}''. Here, P14 began to mind-wander and initiated an inner dialogue about the experiment itself rather than about the VR experience.

\section{Discussion}

Overall, our approach was effective, as we collected 57 BIP episodes. We revealed four main BIP patterns. Participants who experienced \textit{reflected-upon} BIPs reflected on the distractor and then used a strategy to come back in VR; those who experienced \textit{discarded} BIPs noticed the distractor and got back to VR without even thinking about it\footnote{Phenomenologists would say that they experienced these breaks \textit{pre-reflectively}. Micro-phenomenological interviews have, among other things, been designed to help interviewees get back in contact with past pre-reflective experiences, so as to reflect on and verbalize them.}. \textit{Self-preservation} BIPs arose when the fear of heights of participants was so intense that they chose to break presence, while \textit{contradictory mediation} BIPs occurred when participants got aware that an action they wanted to perform was not possible in VR. To our knowledge, our study is the first of its kind to identify and describe the unfolding of BIPs in such detail, as only examples or general descriptions of BIPs were previously available.

We now discuss the BIP episodes and patterns we found with those reported in prior research and propose three design opportunities, before addressing the limitations of our study.

\subsection{Our BIP episodes and the PI/Psi presence model}

We examined whether each identified BIP episode could be related to a break in Place Illusion or a break in Plausibility Illusion.

From the literature on presence, a break in PI can be defined as a break in the user’s subjective sensation of ``being there'' in the virtual environment. We can classify as a break in PI all episodes that included either the top-level generic category \textit{Getting back to reality}, the second-level generic category \textit{Knowing it is not coming from VR} or \textit{Knowing the source is reality}, and synchronic category of \textit{Reality awareness}. Other terms, used by the participant, such as ``coexistence of two distinct worlds'', ``non-virtual reality'' also evoke PI elements. All these elements indicate that participants were no longer experientially present in the virtual environment, their responses being directed toward the real world rather than the virtual one. This resulted in 30 BIP episodes classified as break in Psi.

Concerning break in Psi, it is defined as a break in the illusion that events in the virtual environment are actually happening (even though participants cognitively know that these events are not real)~\cite{slater_place_2009}. According to~\citeauthor{slater_separate_2022}, a break should occurs when either: the virtual environment fails to respond appropriately to the participant’s actions (e.g., an element has not been programmed), when an element of the environment fails to acknowledge participants’ position, or role (e.g. the environment behaves in a purely scripted or generic way), or when the environment fails to meet the participant’s expectations (e.g., objects floating without reason)~\cite{slater_separate_2022}. We can classify as breaks in Psi all episodes that included the top-level diachronic category \textit{Noticing something abnormal}, and the second-level diachronic category \textit{Knowing something about VR}. In the first case, the second-level generic categories \textit{Unfulfilled expectation} and \textit{Sensory incongruence} directly correspond to failures in participants’ expectations or to the failure of \textit{sensorimotor coherence}, as proposed by~\cite{Pouke_2022_qualitative}. In the second case, the category \textit{Knowing something about VR} primarily reflects the understanding that one cannot physically grasp a guardrail in VR. All these led to the classification of 9 BIP episodes as break in Psi.

These results suggest that break in PI can occur without break in Psi, and vice versa, which is consistent with \citeauthor{Pouke_2022_qualitative}’s conclusion that PI and Psi can be independent~\cite{Pouke_2022_qualitative}. Participants also reported being able to recover from a break in Psi, once the source of the incoherence disappeared (e.g., when the two-headed seagull flew away), and by applying a VR-comeback strategy. This is again coherent with recent findings: participants are able to restore Psi once the source of the inconsistency disappeared~\cite{Pouke_2022_qualitative}. 

Interestingly, we also identified 17 BIP episodes that exhibited indicators of both a break in PI and a break in Psi. In all cases, clues of a break in Psi preceded clues of a break in PI. This may indicate that disruptions in the Psi can open the door for reality to intrude on participants' experience, whereas a break in PI may not compromise the overall coherence of the virtual environment. In other words, during a break in Plausibility Illusion, participants may also experience a break in Place Illusion within the same BIP episode.

Note that the BIP episode classified as mind wandering (P14 BIP3) does not fit the definition of either a break in PI or a break in Psi. It may instead correspond to a break in attention, as described in \citeauthor{waterworth_focus_2001}'s framework~\cite{waterworth_focus_2001}, which is a phenomenon that falls outside the PI/Psi framework. Alternatively, this may reflect a break in PI toward a mental imaginary space, as suggested by Biocca’s three-pole model~\cite{biocca2003can}.

Now that we have categorized break episodes as breaks in PI or breaks in Psi, we can examine whether our distractors effectively triggered the intended type of break. The ``music'' distractor, designed to disrupt PI, successfully induced eight breaks in PI and one break in Psi, with one instance of break in Psi occurring within a break in PI. The ``Plank removed'' distractor, intended to provoke both breaks in PI and breaks in Psi, resulted in one break in PI, one break in Psi, and five cases of break in Psi within a break in PI. The ``Door + name'' distractor was intended to provoke a break in PI and successfully did so in 13 instances. It did not trigger any break in Psi, making it the most reliable trigger for disrupting Place Illusion. The ``Black screen'' distractor, also aimed at inducing a break in PI, resulted in two breaks in PI and two cases of break in Psi within a break in PI. Finally, the distractor ``helicopter + tarpaulin + seagulls'', intended to induce cognitive overload and a break in Psi, provoked one break in PI and one break in Psi.

\subsection{Our BIP patterns and the litterature}

Concerning the BIP patterns, \textit{Contradictory mediation} BIPs mostly occurred when participants attempted to grab a virtual guardrail that did not physically exist. This may align with \citeauthor{lee_presence_2004}’s model of presence~\cite{lee_presence_2004}, in which virtual objects are experienced as real, until an interaction exposes their virtual nature. We also found that all \textit{Contradictory mediation} BIPs show clues of, first a break in Psi, and then of a break in PI, suggesting that certain BIP patterns may follow a consistent sequence.

\textit{Reflected-upon} BIPs are more diverse. Some participants described breaking back to reality, which corresponds to a break in PI, whereas the experience of others encompass the diachronic categories of \textit{Sensory incongruence} or \textit{Unfulfilled expectation} which correspond better to a break in Psi. When examining breaks in Psi, we found that they all correspond to \textit{Reflected-upon} BIPs. This suggests that during a break in Psi, participants are more inclined to question the situation and engage in an active inquiry, even if they remain spatially within the virtual world.

\textit{Discarded} BIPs were all classified as breaks in PI, which may indicate, first, that not all breaks in PI are necessarily lived reflectively, and second, that breaks in Psi tend to be meaningful for participants, as none were discarded by participants.

All \textit{Self-preservation} BIPs were classified as break in PI. This is unsurprising, as participants deliberately broke their immersion in order to cross the plank.

Concerning the emotional valence of BIPs, they were predominantly labeled as negative, though not all. We also found that the \textit{Self-preservation} strategy can support users navigate challenging VR experiences, and thus are lived as a relief from VR. Unsurprisingly, no breaks in social presence or embodiment were observed, as the environment was single-player with no avatar.

Lastly, in a seemingly related study, \citeauthor{knibbe_dream_2018} used micro-phenomenology to explore the experience of exiting a VR environment~\cite{knibbe_dream_2018}. Their findings revealed that participants experienced spatial disorientation, changes in their sense of control, heightened awareness of the experimenter, and adaptations to shifts in lighting and texture. These components differ significantly from our findings, suggesting that the experience of exiting VR may not qualify as a BIP.

\subsection{Towards an awareness-based definition of BIPs}

An important result of our study is related to awareness-related synchronic categories, that are consistently present across all BIP episodes: experimentation awareness (45 occurrences), reality awareness (43), and media awareness (26). These moments of awareness were always conscious and could be initiated either by the participants themselves (as seen in \textit{self-preservation} BIPs) or triggered by external distractors. This point relates to the role of attention in presence, as emphasized by~\citeauthor{murphy_what_2020}, who distinguish between two types of attention~\cite{murphy_what_2020}. Perceptual attention occurs when an individual notices something without deliberately allocating mental resources to it, and could corresponds to the behavioral responses some participants showed to distractors, not necessarily indicating a BIP. Cognitive attention, by contrast, refers to voluntary dedicating attention to something. We categorized as BIPs all the episodes in which participants cognitively and consciously reflected on a stimulus (external or internal). The key element in characterizing a BIP may not be the reaction itself, but the conscious awareness of reality, the medium, or experimentation in our case. This may challenge the definition proposed in \citeauthor{slater_physiological_2003}, who describe a BIP as the moment when ``\textit{the participant stops responding to the virtual stream and instead responds to the real sensory stream}''~\cite{slater_physiological_2003}.
In this definition, \citeauthor{slater_physiological_2003} consider voluntary and involuntary changes in behavioral, physiological, and subjective responses as a BIP. However, this definition is driven by actions (voluntary or involuntary), and does not take into account the cognitive processes of becoming aware of something. Instead, we suggest that BIPs can be defined as \textit{the moments when a VR user becomes consciously aware of the reality outside VR or of the medium itself }\footnote{Of course, awareness of the experimental setting is an inherent byproduct of any study and should not be expected outside the lab.}\footnote{Medium can be considered as part of reality, so medium awareness could be subsumed by reality awareness, but we prefer to distinguish between the two because: 1- it corresponds to a difference in the experiences of participants, and 2- it hints toward mediation-based definitions of presence. More work would certainly be needed to further describe the taxonomy of what the VR user may be aware of during a break in presence.}. We also share \citeauthor{murphy_what_2020}’s view that ``\textit{the interplay of attention and presence merits further research}'', particularly in relation to both BIPs and presence~\cite{{murphy_what_2020}}.

Considering BIPs as continuous, durative phenomena (what we called BIP episodes) seems better suited to inform interventions aimed at both preventing the loss of presence and facilitating the return in presence when the participant is ready to re-engage in VR. An \textit{episodic BIP} could then be defined as a meaningful episode of lived-experience around the moment when a VR user becomes consciously aware of either the reality outside the VR or the medium itself, for instance spanning from the moment a distractor is perceived to that of re-immersion.

Finally, some of our participants explained feeling ``in between two worlds'', a phenomenon also observed by \citeauthor{Pouke_2022_qualitative}, in their study on breaks in PI and Psi~\cite{Pouke_2022_qualitative}. This also resonates with the notion of hybrid presence, as when \citeauthor{spagnolli_immersionemersion_2002} demonstrated that individuals could remain focused on the VR environment even during breakdowns, arguing that presence could be distributed over different worlds~\cite{spagnolli_immersionemersion_2002}. In such cases, they added that a sense-making process was needed to turn the disruption into something manageable, which may correspond to the inquiry process we observed in the \textit{reflected-upon} BIP. This suggests that during a break, participants may be aware of both worlds simultaneously, that coexist for the duration of the BIP episode, or of both the VR world and its media-related facticity. 

\subsection{Opportunities for design}
We propose three strategies for minimizing BIPs that stem directly from our results. 

\paragraph{Opportunity 1: Create on-the-fly BIP's sources in VR} 
We found that for some experiences not considered as BIPs, distractors could be integrated into the VR environment (e.g., when P14 explained she believed the music came from a radio within VR). Also, many participants tried to identify the sources of distractors during the \textit{conducting an inquiry} phases of \textit{reflected-upon} BIPs. It should then be possible to try and falsify the answer of such inquiry by creating VR objects on-the-fly, that could be perceived as VR sources of the distractors. For example, if a phone was ringing in the real-world, by adding a virtual phone in VR, from which the actual sound would seem to emanate. Such an approach has already been explored by \citeauthor{tao_integrating_2022} who proposed various methods to integrate haptic, auditory, and olfactory external events~\cite{tao_integrating_2022}. We believe this path should be further explored, focusing for instance on cases when presence is important (e.g., cognitive testing~\cite{ribeiro_VRenv_2024}, meetings, etc.), though a tricky part would be to detect when a distractor should be masked or not, and when to safely keep the user out of reality.

\paragraph{Opportunity 2: Support VR comeback strategies}
We found that participants employed three strategies to re-engage with VR after a BIP: disregarding the distractor, focusing, and returning to action. From a Gibsonian perspective~\cite{gibson_ecological_1979}, human beings are engaged in the world that supports their actions. When this engagement is disrupted in VR, users actively seek to re-establish it, for example through movement which is a way to restore the optical flow. In doing so, they can re-enact a coherent world of action. From these three strategies, three complementary opportunities appear. First, we could facilitate disregarding a non-controlled distractor by introducing another distractor within the VR environment, for instance, sending a bird, or having an avatar directly speak to users, to momentarily capture their attention and help them get back to VR. Second, we could help participants focus on the VR environment by either ensuring that there is always something sufficiently visually attractive to see, or subtly making elements more salient. Third, we could support users to get back to action by making existing call to action more visible, like in-game notifications, or use arrows on the floor to encourage resuming movements which would change the optical flow, and help re-engage. Such passive (always available) or active (generated on the fly) comeback cues would be helpful for any user / VR application.

\paragraph{Opportunity 3: Train users to favor discarded BIPs}
We also found that some BIPs were dismissed without inquiry, with participants not even perceiving distractors as leading to an abnormality (\textit{Noticing something abnormal}\footnote{Internal abnormalities are context-dependent~\cite{slater_separate_2022}; for example, our two-headed seagull would unlikely trigger a BIP on an alien planet.} generic diachronic moment). We believe that such BIPs, even when accompanied by a momentary awareness of reality, are less disruptive than those that are consciously reflected upon. Moreover, in earlier research, \citeauthor{witmer_measuring_1998} hypothesized that users who are willing or able to focus on VR stimuli while ignoring external distractions are likely to be more present~\cite{witmer_measuring_1998}. Therefore, we propose training users to ``manage'' BIPs by helping them become experts at disregarding distractors without conscious effort, which could enable them to experience more BIPs pre-reflectively, as \textit{discarded}.
Reducing elements of surprise could be done during tutorials, by informing users about potential external distractors (saying ``\textit{you may hear somebody passing in the corridor, a car honking}''), training them to recognize external distractors, or showing them some media-related abnormalities likely to occur (e.g., glitches). This would accelerate the inquiry phase for \textit{reflected-upon} BIPs or even lead to developing an expertise at discarding distractors without further thinking. This would especially be useful for non-expert users, e.g., VR novices, who still need to learn how to casually discard distractors and get back to VR.

\subsection{Limitations and future work}

There are limitations related to our work. First, though we pre-tested our distractors, some did not work as expected, such as the ``information overload'' one. This may be due to insufficient recovery time between distractors within the experimental procedure.

Second, we introduced Place and Plausibility Illusion distractors at different moments and locations, doing our best to vary their types, and used a presence inducing environment. We also collected descriptions of various BIPs we did not induce. While the breaks themselves indicate the existence of presence beforehand, we did not systematically assess it. Studying the impact of consecutive BIPs on presence and recovery would be interesting. More generally, it appears, that there is no consensus in the literature regarding the most effective types of distractors for inducing BIPs, the optimal spacing between them, or the best environments to do so. We encourage future research to participate in building shared robust protocols for the study of BIPs by elicitating a structured taxonomy of distractors in relation to the types of BIP they can induce, including guidelines on their design, implementation, and spacing, associated VR environments, ways to asses presence between them, etc.

Third, our novel method to collect phenomenological descriptions of BIPs interestingly unveiled that a lot of participants had frequent awareness of the experimental settings (experimentation awareness). These would by definition not have appeared in ecological contexts, and it might be interesting to conduct the same experiment with VR daily users. This would be feasible, as micro-phenomenological interviews can be used to study past non-experimental lived situations. Of course, this would need to address recent or startling situations where BIP occurred. 

Fourth, an interesting avenue for future research would be to examine the behavior of VR experts with respect to BIPs. We believe that external distractors play a much smaller role when VR knowledge and experience have been accumulated over a long period of use. In such cases, micro-phenomenology may no longer be the most relevant approach, and methods such as Interpretative Phenomenological Analysis (IPA) may be more appropriate. Indeed, IPA would allow researchers to investigate how experienced VR users interpret and make sense of presence and breaks, drawing on multiple past experiences to articulate what these concepts mean to them. Notably, IPA has recently been applied in presence research, highlighting its potential to uncover themes that are often overlooked in  presence literature~\cite{Kelly2022}.

Other limitations deal with our analysis and modeling. First, we had difficulty to organize some moments when arranging within the diachronic structures, especially those belonging to the \textit{Reacting} category, which remain uncertain regarding exact duration or position. 
Systematically collecting and using behavioral data (e.g., head or limb trajectories, or external videos) along experiential data would contribute to the automated detection of BIPs\footnote{For example, detecting head movements unrelated to the environment and the task (stops, turning the head, etc.).}, but also to gain a better understanding of the onset, offset, and duration of participants' reactions during diachronic analysis. Second, we extracted the patterns by organizing episodes according to our generic diachronic moments, with roughly the same beginning, ending, and comparable elements in the middle, including their sequential organization (cf. Table~\ref{tab:BIP}). As it gave interesting patterns, such methodology showed its value to study BIPs, but it may be acceptable only for an exploratory study. Achieving more precise and well-founded conclusions would need further research and more data. For example, one could consider replicating our experiment, with the aim to validate both our diachronic moments and the identified patterns using a confirmatory approach. Also, an interesting topic to focus on could be \textit{reflected-upon} and \textit{discarded} BIPs, so as to check and precise their respective unfolding and better understand how they differ. 

Overall, the approach we propose could pave the way for the development of the comprehensive BIP taxonomy we advocate for the future. Achieving such a goal would also need to aim at the phenomenological description of other types of BIPs, such as those that may be related to embodiment or avatars, as well as, studying BIPs in ecological context. Lastly, let us note that micro-phenomenology could be useful to study many XR-related phenomenon, such as presence itself, embodiment, co-presence, cybersickness, etc.

\section{Conclusion}

We argue that, beyond classifying breaks in presence by their causes or attempting to detect them automatically, it is crucial to deepen our understanding by examining how users actually experience them. In this work, we emphasized that BIPs, as experienced, are not all-or-nothing processes, but rather continuous processes that unfold over time and involve users consciously navigating their way through them. We identified four distinct phenomenological patterns of BIP episodes: \textit{reflected-upon}, \textit{discarded}, \textit{self-preservation}, and \textit{contradictory mediation} ones. We proposed some awareness definitions for momentary or episodic BIPs, and outlined three BIP-related design opportunities based on creating on-the-fly sources of distractors in VR, supporting VR comeback strategies, and favoring discarded BIPs by user training. Our micro-phenomenological study provided valuable insights into the lived experience of BIPs. However, this work is far from complete, and to continue developing a phenomenological model of BIPs that is as accurate as possible, future research should investigate other types of BIPs, aim to establish shared methodologies, and examine them in ecological contexts.

\backmatter

\bmhead{Supplementary information}

This article is accompanied with supplementary files (Online resources). HTML files showcasing our main results can be found in the ``Analysis'' folder. These HTML files correspond to spreadsheets created during our analysis. We standardized the naming convention for each BIP as follow: P\# BIP\# ([Name of the BIP]), where P\# represents the participant ID and BIP\# indicates the order of the BIP within the interview session.

There are five files:
\begin{enumerate}
 \item BIP{\_}patterns.html: Corresponds to the final stage of the generic diachronic analysis (step 4, part 3.6 in the Data Analysis section of the paper), where different BIPs are aligned and grouped into patterns. You will find the 4 patterns we identified (Reflected-upon, Contradictory mediation, Discarded, Self-preservation BIPs).
 \item Diachronic{\_}and{\_}synchronic{\_}analysis.html: Contains the raw data for each BIP. You will find the top-level generic diachronic moments, the second-level generic diachronic moments, the initial diachronic moments, and their corresponding synchronic categories, properties, and property values. Verbatim corresponding to each diachronic moment, synchronic category, and property are not provided for anonymization reasons.

 \item BIP{\_}episode{\_}statistics.html: Corresponds to Table 3 in the paper, and presents some BIP statistics.

\item Distractors{\_}and{\_}types{\_}of{\_}BIPs.html: Corresponds to each of our distractors, the types of BIP found.

\item Types{\_}of{\_}BIP.html: Corresponds to BIP episodes classified as break in PI, break in Psi, or other.

\end{enumerate}

\bmhead{Acknowledgements}
We thank Pierre Vaslin for his efforts in maintaining the virtual environment. We thank all the participants who took part in the study, as well as the pilot testers. The comments of anonymous reviewers largely helped improve its structure and clarity.
This work is part of the RUPTUR project funded by the Pays de la Loire French region through the PULSAR program.

\section*{Declarations}

\begin{itemize}
\item Conflict of interest/Competing interests: The authors declare that they have no known competing financial interests or personal relationships that could have appeared to influence the work reported in this paper.
\item Ethics approval and consent to participate: The protocol was validated by the ethics committee of our university\footnote{CEDIS: Comité d'Ethique, de Déontologie et d'Intégrité Scientifique de Nantes Université IRB number: IORG0011023 - reference number: 22052024-1)}. All participants signed a consent form before the study. 
\item All data is available upon request.
\item Author contribution: All authors contributed to the study conception and design. Material preparation and data collection were performed by Sarah Varlin Grassi, Jean-Philippe Rivière, and Roman Malo. All authors contributed to data analysis, to the first draft of the manuscript, and commented on previous versions of the manuscript. The final draft of the paper was written by Jean-Philippe Rivière and Yannick Prié. All authors read and approved the final manuscript.
\end{itemize}

\newpage
\begin{appendices}

\appendix\section{Result of the SP-IE questionnaire}\label{app:secA1}

\begin{figure}[h!] 
 \hspace{-3cm}
 \includegraphics[width=1.3\linewidth]{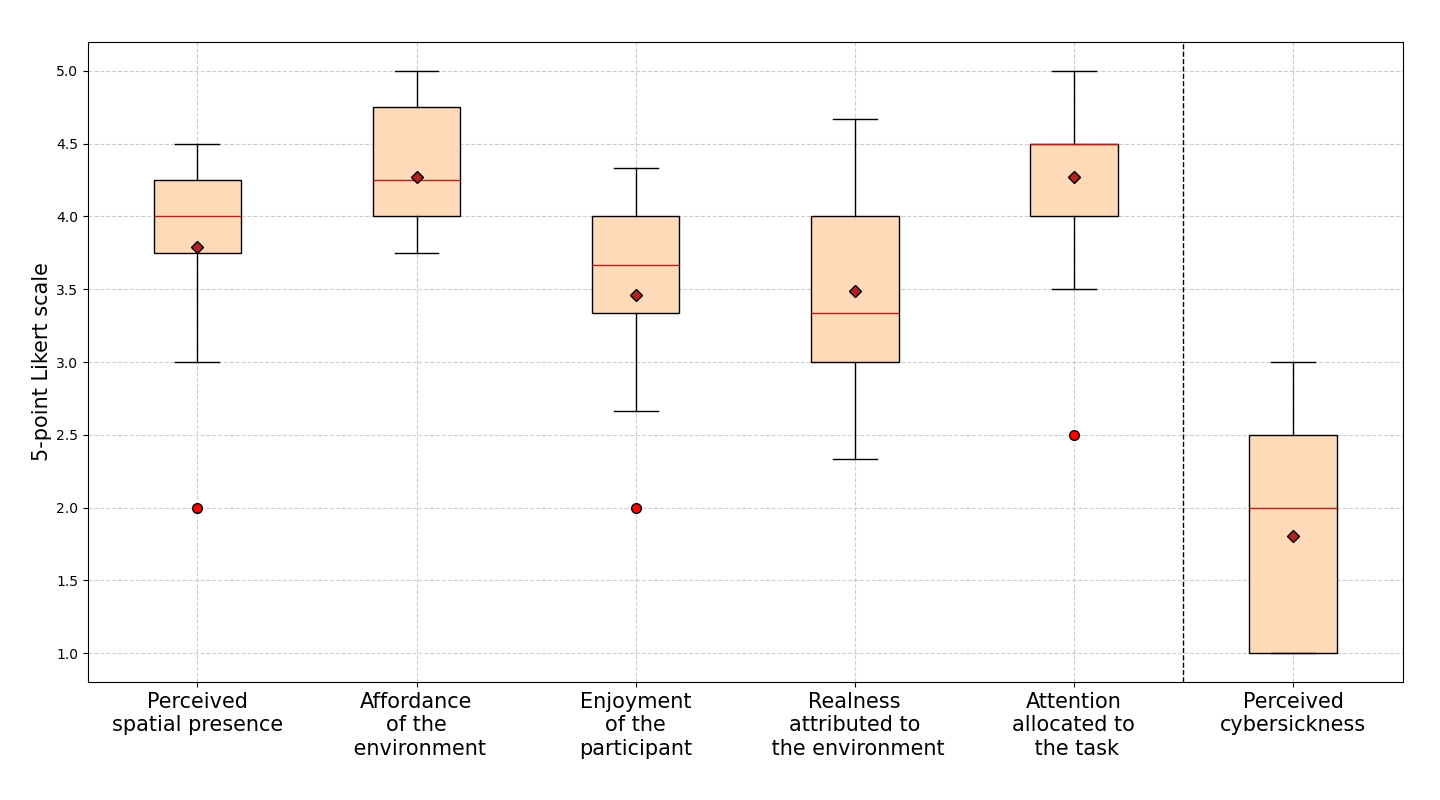}
 \caption{Results for the SP-IE questionnaire with values for each subscale. Red dots indicate outliers, the red bar represents the median, and red diamonds show the means}.
 \label{fig:SP-IE}
 \hfill 
 
\end{figure}
\newpage

\section{Acrophobia assessment questionnaires}\label{secA2}

\input{tables/vhiss}
\newpage

\end{appendices}

\bibliography{references}

\end{document}

%% file: tables/tbl_BIP.tex

\begin{landscape}
\begin{table}[]

\thispagestyle{empty}

\addtolength{\tabcolsep}{-3pt}


\caption{All 57 BIPs classified by source and participant. ``N/A'' values are explained by P5 dropping the session early on, and the impossibility to generate the \textit{black screen} distractor after P6 session}
\label{tab:BIP}
\begin{tabular}{|>{\raggedright}p{0.34\textwidth}|>{\raggedright}p{0.07\textwidth}|>{\raggedright}p{0.07\textwidth}|>{\raggedright}p{0.07\textwidth}|>{\raggedright}p{0.07\textwidth}|>{\raggedright}p{0.07\textwidth}|>{\raggedright}p{0.07\textwidth}|>{\raggedright}p{0.07\textwidth}|>{\raggedright}p{0.07\textwidth}|>{\raggedright}p{0.07\textwidth}|>{\raggedright}p{0.07\textwidth}|>{\raggedright}p{0.07\textwidth}|>{\raggedright}p{0.07\textwidth}|>{\raggedright}p{0.07\textwidth}|>{\raggedright}p{0.07\textwidth}|}

\textbf{Induced BIP}&
  \multicolumn{1}{c}{\textbf{P1}}&
  \multicolumn{1}{c}{\textbf{P2}}&
  \multicolumn{1}{c}{\textbf{P3}}&
  \multicolumn{1}{c}{\textbf{P4}}&
  \multicolumn{1}{c}{\textbf{P5}}&
  \multicolumn{1}{c}{\textbf{P6}}&
  \multicolumn{1}{c}{\textbf{P7}}&
  \multicolumn{1}{c}{\textbf{P8}}&
  \multicolumn{1}{c}{\textbf{P9}}&
  \multicolumn{1}{c}{\textbf{P10}}&
  \multicolumn{1}{c}{\textbf{P11}}&
  \multicolumn{1}{c}{\textbf{P12}}&
  \multicolumn{1}{c}{\textbf{P13}}&
  \multicolumn{1}{c}{\textbf{P14}}\\ \hline
Music &
   &
  \cellcolor[HTML]{CCCCCC}BIP1 &
  \cellcolor[HTML]{CCCCCC}BIP1 &
  \cellcolor[HTML]{CCCCCC}BIP1 &
  N/A&
  \cellcolor[HTML]{CCCCCC}BIP1 &
  \cellcolor[HTML]{CCCCCC}BIP1 &
  \cellcolor[HTML]{CCCCCC}BIP1 &
  \cellcolor[HTML]{CCCCCC}BIP1 &
  \cellcolor[HTML]{CCCCCC}BIP1 &
  \cellcolor[HTML]{CCCCCC}BIP2 &
  \cellcolor[HTML]{CCCCCC}BIP2 &
  \cellcolor[HTML]{CCCCCC}BIP2 &
   \\
Plank Removed &
   &
   &
  &
  \cellcolor[HTML]{CCCCCC}BIP2 &
N/A&
  \cellcolor[HTML]{CCCCCC}BIP3 &
  \cellcolor[HTML]{CCCCCC}BIP3 &
   &
  \cellcolor[HTML]{CCCCCC}BIP2 &
  \cellcolor[HTML]{CCCCCC}BIP2 &
   &
  \cellcolor[HTML]{CCCCCC}BIP3 &
  \cellcolor[HTML]{CCCCCC}BIP3 &
   \\
Door + Name &
  \cellcolor[HTML]{CCCCCC}BIP1&
  \cellcolor[HTML]{CCCCCC}BIP2&
  \cellcolor[HTML]{CCCCCC}BIP3&
  \cellcolor[HTML]{CCCCCC}BIP3&
 N/A&
  \cellcolor[HTML]{CCCCCC}BIP4&
  \cellcolor[HTML]{CCCCCC}BIP4&
  \cellcolor[HTML]{CCCCCC}BIP2&
  \cellcolor[HTML]{CCCCCC}BIP4&
  \cellcolor[HTML]{CCCCCC}BIP3&
  \cellcolor[HTML]{CCCCCC}BIP4&
  \cellcolor[HTML]{CCCCCC}BIP5&
  \cellcolor[HTML]{CCCCCC}BIP4&
  \cellcolor[HTML]{CCCCCC}BIP2\\
Black screens &
  \cellcolor[HTML]{D9D9D9}BIP2&
   &
  \cellcolor[HTML]{CCCCCC}BIP5 &
  \cellcolor[HTML]{CCCCCC}BIP5 &
  N/A&
  \cellcolor[HTML]{CCCCCC}BIP6 &
  N/A&
  N/A&
  N/A&
  N/A&
  N/A&
  N/A&
  N/A&
  N/A\\
  Helicopter + tarpaulin + seagulls
 &
   &
   &
   &
  \cellcolor[HTML]{CCCCCC}BIP4 & N/A
   &
   &
   & 
   &
   &
  \cellcolor[HTML]{CCCCCC}BIP5&
   &
   &
   &
   \\ 
   &
   &
   & 
   & 
   & 
   &
   &
   &
   &
   &
   &
   &
   &
   &
  \textbf{}\\  
\textbf{Not induced BIP}&
  \multicolumn{1}{c}{\textbf{P1}}&
  \multicolumn{1}{c}{\textbf{P2}}&
  \multicolumn{1}{c}{\textbf{P3}}&
  \multicolumn{1}{c}{\textbf{P4}}&
  \multicolumn{1}{c}{\textbf{P5}}&
  \multicolumn{1}{c}{\textbf{P6}}&
  \multicolumn{1}{c}{\textbf{P7}}&
  \multicolumn{1}{c}{\textbf{P8}}&
  \multicolumn{1}{c}{\textbf{P9}}&
  \multicolumn{1}{c}{\textbf{P10}}&
  \multicolumn{1}{c}{\textbf{P11}}&
  \multicolumn{1}{c}{\textbf{P12}}&
  \multicolumn{1}{c}{\textbf{P13}}&
  \multicolumn{1}{c}{\textbf{P14}}\\ \hline Birds (including two-headed seagull)  &
   &
   &
  \cellcolor[HTML]{CCCCCC}BIP4 &
   &
   &
  \cellcolor[HTML]{CCCCCC}BIP5 &
  \cellcolor[HTML]{CCCCCC}BIP5 &
   &
   &
  \cellcolor[HTML]{CCCCCC}BIP4 &
   &
  \cellcolor[HTML]{CCCCCC}BIP6 &
   &
   \\
Crossing the plank &
   &
   &
  \cellcolor[HTML]{CCCCCC}BIP2 &
   &
  \cellcolor[HTML]{CCCCCC}BIP2 &
   &
   &
   &
   &
   &
  \cellcolor[HTML]{CCCCCC}BIP1 &
   &
   &
   \\
The guardrail &
   &
   &
   &
   &
  \cellcolor[HTML]{CCCCCC}BIP1 &
  \cellcolor[HTML]{CCCCCC}BIP2 &
   &
   &
   &
   &
   &
  \cellcolor[HTML]{CCCCCC}BIP1 &
  \cellcolor[HTML]{CCCCCC}BIP1 &
   \\
Block of concrete &
   &
   &
   &
   &
   &
   &
  \cellcolor[HTML]{CCCCCC}BIP2 &
   &
   &
   &
   &
   &
   &
   \\
Phone notification &
   &
   &
   &
   &
   &
   &
   &
   &
  \cellcolor[HTML]{CCCCCC}BIP3 &
   &
   &
   &
   &
   \\
Tarpaulin &
   &
   &
   &
   &
   &
   &
   &
   &
   &
  \cellcolor[HTML]{CCCCCC}BIP6 &
   &
   &
   &
   \\
The metal plank &
   &
   &
   &
   &
   &
   &
   &
   &
   &
   &
  \cellcolor[HTML]{CCCCCC}BIP3 &
  \cellcolor[HTML]{CCCCCC}BIP4 &
   &
  \cellcolor[HTML]{CCCCCC}BIP1 \\
  Seeing the experimentation room 
&
   &
   &
   &
   &
   &
   &
   &
   &
   &
   &
  \cellcolor[HTML]{CCCCCC}BIP5  &
   &
   &
   \\
Mind wandering &
   &
   &
   &
   &
   &
   &
   &
   &
   &
   &
   &
   &
   &
  \cellcolor[HTML]{CCCCCC}BIP3 \\ 
\end{tabular}%

\end{table}
\end{landscape}

%% file: tables/tbl_BIPRU.tex
\begin{table}[]

\caption{Five examples of Reflected-upon BIPs presented with generic diachronic moments. Reflected-upon BIPs are characterized by the presence of at least one diachronic category: \textit{Something abnormal} (turquoise), \textit{Conducting an inquiry} (purple), or an understanding of what is happening (pink). They usually finish with a strategy to return in VR (blue)}
\label{tab:BIPRU}
\begin{tabular}{ll|l|llllll}
\multicolumn{1}{|c|}{\cellcolor[HTML]{CCCCCC}\begin{tabular}[c]{@{}c@{}}P1 BIP1\\ (Door + name)\end{tabular}} &  & \multicolumn{1}{c|}{\cellcolor[HTML]{CCCCCC}\begin{tabular}[c]{@{}c@{}}P12 BIP5\\ (Door + name)\end{tabular}} & \multicolumn{1}{l|}{} & \multicolumn{1}{c|}{\cellcolor[HTML]{CCCCCC}\begin{tabular}[c]{@{}c@{}}P7 BIP1\\ (Music)\end{tabular}} & \multicolumn{1}{l|}{} & \multicolumn{1}{c|}{\cellcolor[HTML]{CCCCCC}\begin{tabular}[c]{@{}c@{}}P13 BIP2\\ (Music)\end{tabular}} & \multicolumn{1}{l|}{} & \multicolumn{1}{c|}{\cellcolor[HTML]{CCCCCC}\begin{tabular}[c]{@{}c@{}}P9 BIP2\\ (Plank removed)\end{tabular}} \\ 
\multicolumn{1}{|l|}{\cellcolor[HTML]{FFF2CC}\begin{tabular}[c]{@{}l@{}}Hearing\\ something\end{tabular}} &  & \cellcolor[HTML]{FFF2CC}\begin{tabular}[c]{@{}l@{}}Hearing\\ something\end{tabular} & \multicolumn{1}{l|}{} & \multicolumn{1}{l|}{\cellcolor[HTML]{FFF2CC}\begin{tabular}[c]{@{}l@{}}Hearing\\ something\end{tabular}} & \multicolumn{1}{l|}{} & \multicolumn{1}{l|}{\cellcolor[HTML]{FFE599}\begin{tabular}[c]{@{}l@{}}Seeing\\ something\end{tabular}} & \multicolumn{1}{l|}{} & \multicolumn{1}{l|}{\cellcolor[HTML]{FFE599}\begin{tabular}[c]{@{}l@{}}Seeing\\ something\end{tabular}} \\
\multicolumn{1}{|l|}{} &  &  & \multicolumn{1}{l|}{} & \multicolumn{1}{l|}{\cellcolor[HTML]{FF6F6F}Stopping} & \multicolumn{1}{l|}{} & \multicolumn{1}{l|}{\cellcolor[HTML]{FF6F6F}Stopping} & \multicolumn{1}{l|}{} & \multicolumn{1}{l|}{\cellcolor[HTML]{00B0A1}\begin{tabular}[c]{@{}l@{}}Sensory\\ incongruence\end{tabular}} \\
\multicolumn{1}{|l|}{\cellcolor[HTML]{D9D2E9}\begin{tabular}[c]{@{}l@{}}Making hypothesis\\  about something\end{tabular}} &  & \cellcolor[HTML]{351C75}{\color[HTML]{FFFFFF} \begin{tabular}[c]{@{}l@{}}Wondering about\\  the source of\\  something\end{tabular}} & \multicolumn{1}{l|}{} & \multicolumn{1}{l|}{\cellcolor[HTML]{D9D2E9}\begin{tabular}[c]{@{}l@{}}Making hypothesis \\ about something\end{tabular}} & \multicolumn{1}{l|}{} & \multicolumn{1}{l|}{\cellcolor[HTML]{674EA7}{\color[HTML]{FFFFFF} \begin{tabular}[c]{@{}l@{}}Testing\\ hypothesis\end{tabular}}} & \multicolumn{1}{l|}{} & \multicolumn{1}{l|}{\cellcolor[HTML]{8E7CC3}\begin{tabular}[c]{@{}l@{}}Analysing\\ the surrounding\end{tabular}} \\
\multicolumn{1}{|l|}{\cellcolor[HTML]{351C75}{\color[HTML]{FFFFFF} \begin{tabular}[c]{@{}l@{}}Wondering about\\  the source of \\ something\end{tabular}}} &  & \cellcolor[HTML]{8E7CC3}\begin{tabular}[c]{@{}l@{}}Analysing \\ the surrounding\end{tabular} & \multicolumn{1}{l|}{} & \multicolumn{1}{l|}{\cellcolor[HTML]{8E7CC3}\begin{tabular}[c]{@{}l@{}}Analysing\\ the surrounding\end{tabular}} & \multicolumn{1}{l|}{} & \multicolumn{1}{l|}{} & \multicolumn{1}{l|}{} & \multicolumn{1}{l|}{} \\
\multicolumn{1}{|l|}{} &  &  & \multicolumn{1}{l|}{} & \multicolumn{1}{l|}{\cellcolor[HTML]{674EA7}{\color[HTML]{FFFFFF} \begin{tabular}[c]{@{}l@{}}Testing\\ hypothesis\end{tabular}}} & \multicolumn{1}{l|}{} & \multicolumn{1}{l|}{} & \multicolumn{1}{l|}{} & \multicolumn{1}{l|}{} \\
\multicolumn{1}{|l|}{\cellcolor[HTML]{F0BED4}\begin{tabular}[c]{@{}l@{}}Knowing the \\ source is reality\end{tabular}} &  & \cellcolor[HTML]{D5A6BD}\begin{tabular}[c]{@{}l@{}}Knowing it is \\ not coming \\ from VR\end{tabular} & \multicolumn{1}{l|}{} & \multicolumn{1}{l|}{\cellcolor[HTML]{F0BED4}\begin{tabular}[c]{@{}l@{}}Knowing the \\ source is reality\end{tabular}} & \multicolumn{1}{l|}{} & \multicolumn{1}{l|}{\cellcolor[HTML]{D5A6BD}\begin{tabular}[c]{@{}l@{}}Knowing it is not \\ coming from VR\end{tabular}} & \multicolumn{1}{l|}{} & \multicolumn{1}{l|}{} \\
\multicolumn{1}{|l|}{} &  &   & \multicolumn{1}{l|}{} & \multicolumn{1}{l|}{} & \multicolumn{1}{l|}{} & \multicolumn{1}{l|}{\cellcolor[HTML]{CC0000}Other reaction} & \multicolumn{1}{l|}{} & \multicolumn{1}{l|}{} \\
\multicolumn{1}{|l|}{} &  & \cellcolor[HTML]{C9DAF8}\begin{tabular}[c]{@{}l@{}}Disregarding \\ the distractor\end{tabular} & \multicolumn{1}{l|}{} & \multicolumn{1}{l|}{\cellcolor[HTML]{A4C2F4}Focusing} & \multicolumn{1}{l|}{} & \multicolumn{1}{l|}{\cellcolor[HTML]{A4C2F4}Focusing} & \multicolumn{1}{l|}{} & \multicolumn{1}{l|}{\cellcolor[HTML]{A4C2F4}Focusing} \\ 
\multicolumn{1}{|l|}{} &  &  & \multicolumn{1}{l|}{} & \multicolumn{1}{l|}{\cellcolor[HTML]{C9DAF8}\begin{tabular}[c]{@{}l@{}}Disregarding \\ the distractor\end{tabular}} &  &  & &  \\
\multicolumn{1}{|l|}{} &  & \cellcolor[HTML]{ffffff}\begin{tabular}[c]{@{}l@{}} ~~~\\~~~ \end{tabular}  & \multicolumn{1}{l|}{} & \multicolumn{1}{l|}{\cellcolor[HTML]{6D9EEB}Back in action} &  &  &  &  \\ 
\multicolumn{1}{|l|}{\cellcolor[HTML]{FFF2CC}\begin{tabular}[c]{@{}l@{}}Hearing\\  something\end{tabular}} &  & \cellcolor[HTML]{FFF2CC}\begin{tabular}[c]{@{}l@{}}Hearing\\  something\end{tabular} &  &  &  &  &  &  \\
\multicolumn{1}{|l|}{\cellcolor[HTML]{D9D2E9}\begin{tabular}[c]{@{}l@{}}Making hypothesis\\  about something\end{tabular}} &  &  &  &  &  &  &  &  \\
\multicolumn{1}{|l|}{\cellcolor[HTML]{C27BA0}\begin{tabular}[c]{@{}l@{}}Knowing something\\ about reality\end{tabular}} &  & \cellcolor[HTML]{F0BED4}\begin{tabular}[c]{@{}l@{}}Knowing the \\ source is reality\end{tabular} &  &  &  &  &  &  \\
\multicolumn{1}{|l|}{\cellcolor[HTML]{FF6F6F}Stopping} &  &  &  &  &  &  &  &  \\
\multicolumn{1}{|l|}{\cellcolor[HTML]{6D9EEB}Back in action} &  & \cellcolor[HTML]{C9DAF8}\begin{tabular}[c]{@{}l@{}}Disregarding \\ the distractor\end{tabular} &  &  &  &  &  &  \\ 
\cellcolor[HTML]{ffffff}\begin{tabular}[c]{@{}l@{}} ~~\\~~~ \end{tabular} &  & \cellcolor[HTML]{6D9EEB}Back in action&  &  &  &  &  &  \\ 
\end{tabular}
\end{table}

%% file: tables/synchronic.tex
\begin{table}[!h]
    \centering
    \caption{The main synchronic categories ranked by number of occurrences and participants who reported them.}
    \begin{tabular}{|l|c|r|}
    \hline
        \textbf{Synchronic categories} & \textbf{\# occurences} & \textbf{\# of participants}  \\ \hline
        Reality awareness & 46  & 14/14 \\ \hline
        Experimentation awareness & 45  & 13/14 \\ \hline
        Attentional focusing & 29  & 12/14 \\ \hline
        Media awareness & 29  & 12/14 \\ \hline
        Abnormality & 23 & 9/14 \\ \hline
        Judgment & 22  & 12/14 \\ \hline
        Fear & 15  & 6/14 \\ \hline
        Surprise & 14 & 8/14 \\ \hline
        Volition & 14 & 8/14 \\ \hline
        Attention & 13 & 7/14 \\ \hline
        Questionning & 11 & 5/14 \\ \hline
        Weirdness & 10  & 7/14 \\ \hline
        Bothered & 10  & 6/14 \\ \hline
        Reassurance & 9 & 7/14 \\ \hline
        Uncomfortable & 9  & 7/14 \\ \hline
    \end{tabular}
    \label{tab:synchronic}
\end{table}

%% file: tables/tbl_BIPD.tex
\begin{table}[]
\caption{Four examples of Discarded BIPs. Discarded BIPs are mainly characterized by the generic diachronic categories \textit{Perceiving something} (yellow) and \textit{Implementing a VR comeback strategy} (blue). 
}
\label{tab:BIPD}

\begin{tabular}{llllll|l|}
\multicolumn{1}{|c|}{\cellcolor[HTML]{CCCCCC}\begin{tabular}[c]{@{}c@{}}P3 BIP5\\ (Black screens)\end{tabular}} &
  \multicolumn{1}{l|}{} &
  \multicolumn{1}{c|}{\cellcolor[HTML]{CCCCCC}\begin{tabular}[c]{@{}c@{}}P4 BIP3\\ (Door + name)\end{tabular}} &
  \multicolumn{1}{l|}{} &
  \multicolumn{1}{c|}{\cellcolor[HTML]{CCCCCC}\begin{tabular}[c]{@{}c@{}}P8 BIP2\\ (Door + name)\end{tabular}} &
   &
  \multicolumn{1}{c|}{\cellcolor[HTML]{CCCCCC}\begin{tabular}[c]{@{}c@{}}P8 BIP1 \\ (Music)\end{tabular}} \\
\multicolumn{1}{|l|}{\cellcolor[HTML]{FFE599}Seeing something} &
  \multicolumn{1}{l|}{} &
  \multicolumn{1}{l|}{\cellcolor[HTML]{FFF2CC}Hearing something} &
  \multicolumn{1}{l|}{} &
  \multicolumn{1}{l|}{\cellcolor[HTML]{FFF2CC}Hearing something} &
   &
  \cellcolor[HTML]{FFF2CC}Hearing something \\
\multicolumn{1}{|l|}{\cellcolor[HTML]{FFE599}Seeing something} &
  \multicolumn{1}{l|}{} &
  \multicolumn{1}{l|}{\cellcolor[HTML]{FFF2CC}Hearing something} &
  \multicolumn{1}{l|}{} &
  \multicolumn{1}{l|}{\cellcolor[HTML]{FFF2CC}Hearing something} &
   &
   \\
\multicolumn{1}{|l|}{\cellcolor[HTML]{8EFC67}Being less in VR} &
  \multicolumn{1}{l|}{} &
  \multicolumn{1}{l|}{\cellcolor[HTML]{8EFC67}Being less in VR} &
  \multicolumn{1}{l|}{} &
  \multicolumn{1}{l|}{} &
   &
   \\
\multicolumn{1}{|l|}{} & 
  \multicolumn{1}{l|}{} &
  \multicolumn{1}{l|}{} &
  \multicolumn{1}{l|}{} &
  \multicolumn{1}{l|}{} &
   &
  \cellcolor[HTML]{FF6F6F}Stopping \\
\multicolumn{1}{|l|}{} &
  \multicolumn{1}{l|}{} &
  \multicolumn{1}{l|}{} &
  \multicolumn{1}{l|}{} &
  \multicolumn{1}{l|}{} &
   &
   \\
\multicolumn{1}{|l|}{\cellcolor[HTML]{C9DAF8}\begin{tabular}[c]{@{}l@{}}Disregarding \\ the distractor\end{tabular}} &
  \multicolumn{1}{l|}{} &
  \multicolumn{1}{l|}{\cellcolor[HTML]{C9DAF8}\begin{tabular}[c]{@{}l@{}}Disregarding \\ the distractor\end{tabular}} &
  \multicolumn{1}{l|}{} &
  \multicolumn{1}{l|}{\cellcolor[HTML]{6D9EEB}Back in action} &
   &
  \cellcolor[HTML]{C9DAF8}\begin{tabular}[c]{@{}l@{}}Disregarding \\ the distractor\end{tabular} \\ 
 &
   &
   &
   &
   &
   &
  \cellcolor[HTML]{C9DAF8}\begin{tabular}[c]{@{}l@{}}Disregarding \\ the distractor\end{tabular}\\ 
\end{tabular}%

\end{table}

%% file: tables/tbl_BIPSD.tex

\begin{table}[]
\caption{Five examples of self-preservation BIPs. They are characterized by the fact that participants deliberately triggered them to manage their fear of heights ---see the generic diachronic moment \textit{Telling oneself it is not real} (orange)--- and the absence of strategies to come back in VR, unlike other BIP patterns.}
\label{tab:BIPSD}
\begin{tabular}{llll|l|llll}


\multicolumn{1}{|c|}{\cellcolor[HTML]{CCCCCC}\begin{tabular}[c]{@{}c@{}}P7 BIP3\\ (Plank removed)\end{tabular}} &
  \multicolumn{1}{c|}{} &
  \multicolumn{1}{c|}{\cellcolor[HTML]{CCCCCC}\begin{tabular}[c]{@{}c@{}}P3 BIP2 \\ (Crossing \\the plank)\end{tabular}} &
   &
  \multicolumn{1}{c|}{\cellcolor[HTML]{CCCCCC}\begin{tabular}[c]{@{}c@{}}P10 BIP2\\ (Plank removed)\end{tabular}} &
  \multicolumn{1}{l|}{} &
  \multicolumn{1}{c|}{\cellcolor[HTML]{CCCCCC}\begin{tabular}[c]{@{}c@{}}P5 BIP1\\ (Crossing \\the plank)\end{tabular}} &
  \multicolumn{1}{l|}{} &
  \multicolumn{1}{c|}{\cellcolor[HTML]{CCCCCC}\begin{tabular}[c]{@{}c@{}}P11 BIP1\\ (Crossing \\the plank)\end{tabular}} \\
\multicolumn{1}{|r|}{} &
  \multicolumn{1}{l|}{} &
  \multicolumn{1}{l|}{} &
   &
   &
  \multicolumn{1}{l|}{} &
  \multicolumn{1}{l|}{\cellcolor[HTML]{FFD966}Feeling something} &
  \multicolumn{1}{l|}{} &
  \multicolumn{1}{l|}{\cellcolor[HTML]{FFE599}Seeing something} \\
\multicolumn{1}{|l|}{} &
  \multicolumn{1}{l|}{} &
  \multicolumn{1}{l|}{} &
   &
   &
  \multicolumn{1}{l|}{} &
  \multicolumn{1}{l|}{} &
  \multicolumn{1}{l|}{} &
  \multicolumn{1}{l|}{} \\
\multicolumn{1}{|l|}{} &
  \multicolumn{1}{l|}{} &
  \multicolumn{1}{l|}{} &
   &
   &
  \multicolumn{1}{l|}{} &
  \multicolumn{1}{l|}{} &
  \multicolumn{1}{l|}{} &
  \multicolumn{1}{l|}{} \\
\multicolumn{1}{|l|}{\cellcolor[HTML]{00DCC9}\begin{tabular}[c]{@{}l@{}} Unfulfilled \\expectation \end{tabular}} &
  \multicolumn{1}{l|}{} &
  \multicolumn{1}{l|}{\cellcolor[HTML]{00DCC9}\begin{tabular}[c]{@{}l@{}} Unfulfilled \\expectation   \end{tabular}} &
   &
  \cellcolor[HTML]{00DCC9}\begin{tabular}[c]{@{}l@{}} Unfulfilled \\expectation   \end{tabular} &
  \multicolumn{1}{l|}{} &
  \multicolumn{1}{l|}{} &
  \multicolumn{1}{l|}{} &
  \multicolumn{1}{l|}{} \\
\multicolumn{1}{|l|}{\cellcolor[HTML]{D9D2E9}\begin{tabular}[c]{@{}l@{}}Making hypothesis\\ about something  \end{tabular}} &
  \multicolumn{1}{l|}{} &
  \multicolumn{1}{l|}{\cellcolor[HTML]{00B0A1}\begin{tabular}[c]{@{}l@{}} Sensory\\ incongruence \end{tabular}} &
   &
  \multicolumn{1}{c|}{}&
  \multicolumn{1}{l|}{} &
  \multicolumn{1}{c|}{{}} &
  \multicolumn{1}{l|}{} &
  \multicolumn{1}{l|}{} \\
\multicolumn{1}{|l|}{\cellcolor[HTML]{674EA7}{\color[HTML]{FFFFFF} Testing hypothesis}} &
  \multicolumn{1}{l|}{} &
  \multicolumn{1}{l|}{} &
   &
  \multicolumn{1}{c|}{} &
  \multicolumn{1}{l|}{} &
  \multicolumn{1}{l|}{} &
  \multicolumn{1}{l|}{} &
  \multicolumn{1}{l|}{} \\
\multicolumn{1}{|l|}{\cellcolor[HTML]{C27BA0}\begin{tabular}[c]{@{}l@{}} Knowing something\\ about reality  \end{tabular}} &
  \multicolumn{1}{l|}{} &
  \multicolumn{1}{l|}{} &
   &
   &
  \multicolumn{1}{l|}{} &
  \multicolumn{1}{l|}{} &
  \multicolumn{1}{l|}{} &
  \multicolumn{1}{l|}{} \\
\multicolumn{1}{|l|}{\cellcolor[HTML]{B4A7D6}Asking questions} &
  \multicolumn{1}{l|}{} &
  \multicolumn{1}{l|}{} &
   &
   &
  \multicolumn{1}{l|}{} &
  \multicolumn{1}{l|}{} &
  \multicolumn{1}{l|}{} &
  \multicolumn{1}{l|}{\cellcolor[HTML]{B4A7D6}Asking questions} \\
\multicolumn{1}{|l|}{} &
  \multicolumn{1}{l|}{} &
  \multicolumn{1}{l|}{} &
   &
   &
  \multicolumn{1}{l|}{} &
  \multicolumn{1}{l|}{} &
  \multicolumn{1}{l|}{} &
  \multicolumn{1}{l|}{} \\
\multicolumn{1}{|l|}{\cellcolor[HTML]{FF6D01}\begin{tabular}[c]{@{}l@{}}Telling oneself \\it is not real\end{tabular}} &
  \multicolumn{1}{l|}{} &
  \multicolumn{1}{l|}{\cellcolor[HTML]{FF6D01}\begin{tabular}[c]{@{}l@{}}Telling oneself \\it is not real\end{tabular}} &
   &
  \cellcolor[HTML]{FF6D01}\begin{tabular}[c]{@{}l@{}}Telling oneself \\it is not real\end{tabular} &
  \multicolumn{1}{l|}{} &
  \multicolumn{1}{l|}{\cellcolor[HTML]{FF6D01}\begin{tabular}[c]{@{}l@{}}Telling oneself \\it is not real\end{tabular}} &
  \multicolumn{1}{l|}{} &
  \multicolumn{1}{l|}{\cellcolor[HTML]{FF6D01}\begin{tabular}[c]{@{}l@{}}Telling oneself \\it is not real\end{tabular}} \\ 
 &
   &
   &
   &
   &
   &
   &
   &
   \\
 &
   &
   &
   &
  \cellcolor[HTML]{00B0A1}\begin{tabular}[c]{@{}l@{}} Sensory\\ incongruence \end{tabular}&
   &
   &
   &
   \\
 &
   &
   &
   &
  \cellcolor[HTML]{D9D2E9}\begin{tabular}[c]{@{}l@{}}Making hypothesis\\ about something  \end{tabular} &
   &
   &
   &
   \\ 
\end{tabular}%

\end{table}

%% file: tables/vhiss.tex
\begin{table}[h!]
\footnotesize
\caption{Score of the acrophobia questionnaire}
\label{app:vHISSandAQ}
\begin{tabular}{|l|c|c|c|c|c|c|c|c|c|c|c|c|c|c|}
\hline
 & P1 & P2 & P3 & P4 & P5 & P6 & P7 & P8 & P9 & P10 & P11 & P12 & P13 & P14 \\ \hline
AQ-Anxiety & 16 & 3 & 36 & 30 & 44 & 2 & 9 & 15 & 11 & 15 & 30 & 21 & 6 & 33 \\ \hline
vHISS      & 0  & 0 & 0  & 2  & 0  & 0 & 0 & 0  & 0  & 0  & 4  & 0  & 0 & 0  \\ \hline
\end{tabular}
\end{table}